\pdfoutput=1
\RequirePackage{lineno}
\documentclass[preprint2,times,tighten,trackchanges]{aastex6}
\usepackage{lineno} 
\usepackage{amsmath,amssymb}
\usepackage{graphicx}
\usepackage{url}
\usepackage{hyperref}
\usepackage{color}
\usepackage{multirow}
\usepackage{placeins}
\usepackage{float}	
\usepackage{rotating}
\usepackage{xcolor}

\shorttitle{Sun's cosmic ray shadow with IceCube} 
\shortauthors{M.~G.~Aartsen et al.} 	
																
\begin{document}
\title{Detection of the temporal variation of the Sun's cosmic ray shadow with the IceCube detector}
\author{
IceCube Collaboration\textsuperscript{$\dagger$}:
M.~G.~Aartsen\altaffilmark{1},
M.~Ackermann\altaffilmark{2},
J.~Adams\altaffilmark{1},
J.~A.~Aguilar\altaffilmark{3},
M.~Ahlers\altaffilmark{4},
M.~Ahrens\altaffilmark{5},
D.~Altmann\altaffilmark{6},
K.~Andeen\altaffilmark{7},
T.~Anderson\altaffilmark{8},
I.~Ansseau\altaffilmark{3}
% ...
}
\altaffiltext{1}{Dept.~of Physics and Astronomy, University of Canterbury, Private Bag 4800, Christchurch, New Zealand}
\altaffiltext{2}{DESY, D-15738 Zeuthen, Germany}
\altaffiltext{3}{Universit\'e Libre de Bruxelles, Science Faculty CP230, B-1050 Brussels, Belgium}
\altaffiltext{4}{Niels Bohr Institute, University of Copenhagen, DK-2100 Copenhagen, Denmark}
\altaffiltext{5}{Oskar Klein Centre and Dept.~of Physics, Stockholm University, SE-10691 Stockholm, Sweden}
\altaffiltext{6}{Erlangen Centre for Astroparticle Physics, Friedrich-Alexander-Universit\"at Erlangen-N\"urnberg, D-91058 Erlangen, Germany}
\altaffiltext{7}{Department of Physics, Marquette University, Milwaukee, WI, 53201, USA}
\altaffiltext{8}{Dept.~of Physics, Pennsylvania State University, University Park, PA 16802, USA}
% ...
\begin{abstract}
We report on the observation of a deficit in the cosmic ray flux from the directions of the Moon and Sun with five years of data taken by the IceCube Neutrino Observatory. Between May 2010 and May 2011 the IceCube detector operated with 79 strings deployed in the glacial ice at the South Pole, and with 86 strings between May 2011 and May 2015. A binned analysis is used to measure the relative deficit and significance of the cosmic ray shadows. Both the cosmic ray Moon and Sun shadows are detected with high statistical significance ($>10\sigma$) for each year. The results for the Moon shadow are consistent with previous analyses and verify the stability of the IceCube detector over time. This work represents the first observation of the Sun shadow with the IceCube detector. We show that the cosmic ray shadow of the Sun varies with time. These results open the possibility to study cosmic ray transport near the Sun with future data from IceCube.
\end{abstract}

\maketitle
\textsuperscript{$\dagger$}E-mail: analysis@icecube.wisc.edu
\clearpage
\section{Introduction}
The IceCube Neutrino Observatory \citep{icecube2017} comprises a 1\,km$^3$ Cherenkov light detector buried in the Antarctic ice between 1450 and 2450 meters below the geographic South Pole. 
The trigger rate is dominated by atmospheric muons that are produced when cosmic ray particles interact with the Earth's atmosphere. 
Cosmic rays from the direction of the Moon and Sun are absorbed by the celestial bodies. 
The deficit in the cosmic ray flux toward the direction of the Moon was already observed by IceCube with its 40 and 59 string configurations, using data collected from May 2008 to May 2010. 
Two independent analyses observed the deficit in the cosmic ray flux with high statistical significance ($>6\sigma$) \citep{Aartsen:2013zka}.

The main purpose of the previous paper was to verify the pointing of the detector. 
In this paper, we investigate for the first time the Sun shadow as well. 
Here, the goal is to study the temporal behavior of the two shadows: while the Moon is expected to be stable in time (with only smaller variations expected due to the changing apparent size), the Sun shadow has been shown to vary with the solar cycle and, in turn, with the Sun's magnetic field \citep{2013PhRvL.111a1101A} and has been used to measure the mean interplanetary magnetic field \citep{argo_sun}.
It is the goal of this analysis to investigate if the effect of a temporal variation is present even at the high energies that IceCube is sensitive to (median primary particle energy of events passing the Moon and Sun filters: $\sim 40$~TeV, see section \ref{sec:simulation}).

IceCube detects cosmic rays via high-energy secondary muons, which are produced when cosmic ray particles interact with the Earth's atmosphere. 
The Moon and the Sun subtend an angle of $\sim 0.5^{\circ}$, which is smaller than the median angular resolution for muon detection. 
The Moon can be used to estimate the resolution and pointing in IceCube. 
Other experiments also used the Moon as a calibrator to measure the angular resolution and absolute pointing capabilities of their detectors (see \cite{Ambrosio:2003mz}, \cite{Achard:2005az}, \cite{Adamson:2010hv}, \cite{2011PhRvD..84b2003B}, \cite{hawc2013} and \cite{2010APh....33...97O}). 
However, the Sun cannot be used to calibrate the IceCube detector as the solar magnetic field is expected to influence cosmic rays on their way to Earth. 

The Tibet-AS Gamma experiment analyzed the cosmic ray shadow caused by the Sun (median primary particle energy: $\sim 10$~TeV) and discovered an influence of the solar magnetic field, presented in \cite{2013PhRvL.111a1101A}. 
The depth of the Sun shadow varied during the observation period from 1996 through 2009, while the cosmic ray shadow of the Moon was measured to be constant over time within the experimental uncertainties in \cite{2013PhRvL.111a1101A}, as expected. 
Tibet finds that the variation can be well-described by including cosmic ray transport in the solar magnetic field. 

It is the aim of this paper to use IceCube data to provide an independent observation of the temporal variation of the Sun shadow. 
In addition, IceCube measures particles at a mean energy of around $40$~TeV, which is at significantly higher energies as compared to Tibet. 
Establishing the measurement of the cosmic ray shadow of the Sun with IceCube and other experiments will provide long term energy-dependent information on the cosmic ray transport in the solar magnetic field, which makes this study particularly interesting. 

The IceCube cosmic ray Moon and Sun shadow analysis is based on a binned analysis that measures the relative deficit of events from the direction of the celestial bodies. 
Additionally, a two-dimensional visualization of the shadow is derived. 
Both methods make use of data from on- and off-source regions with events taken from the same elevation as the Moon and Sun but at different right ascension. 
This is possible due to IceCube's uniform exposure in right ascension. 

\section{Detector configuration and Data Sample}
\subsection{IceCube Detector}
The IceCube Neutrino Observatory comprises two components: the IceCube neutrino telescope, which detects penetrating muons and neutrinos, and the IceTop surface array, which detects extensive air showers induced by high energy cosmic rays. 
The IceCube neutrino telescope makes use of the glacial ice as a detection medium for the Cherenkov radiation from charged particles. 
Charged relativistic particles emit Cherenkov light in media, where the speed of light is reduced by the index of refraction. 
IceCube consists of 5160 Digital Optical Modules (DOMs) that are arranged on 86 vertical cables (called strings). 
Each DOM contains a 10 inch photomultiplier tube enclosed in a glass pressure sphere. 

IceCube's primary goal is to detect high energy neutrinos from astrophysical sources, see e.g. \cite{2010RScI...81h1101H} and \cite{2008PhR...458..173B} for reviews. 
Cosmic-ray induced muons and neutrinos from the atmosphere, which are a background for astrophysical neutrino searches, are far more abundant and comprise most IceCube triggers. 
These cosmic ray events can be used to investigate particle interaction physics in the forward direction, to study cosmic rays themselves, or to trace back cosmic ray directions as in this work. 
In this analysis, atmospheric muons are used to investigate the shadow in the cosmic ray flux caused by the Sun and the Moon. 
As the detected muons are highly relativistic, the direction of the cosmic ray can be derived from the muon's direction. 
Estimating the average primary cosmic ray energy of the sample is also possible through comparison to simulations. 

\begin{figure}[htbp]
\centering
\includegraphics[width=1.0\columnwidth]{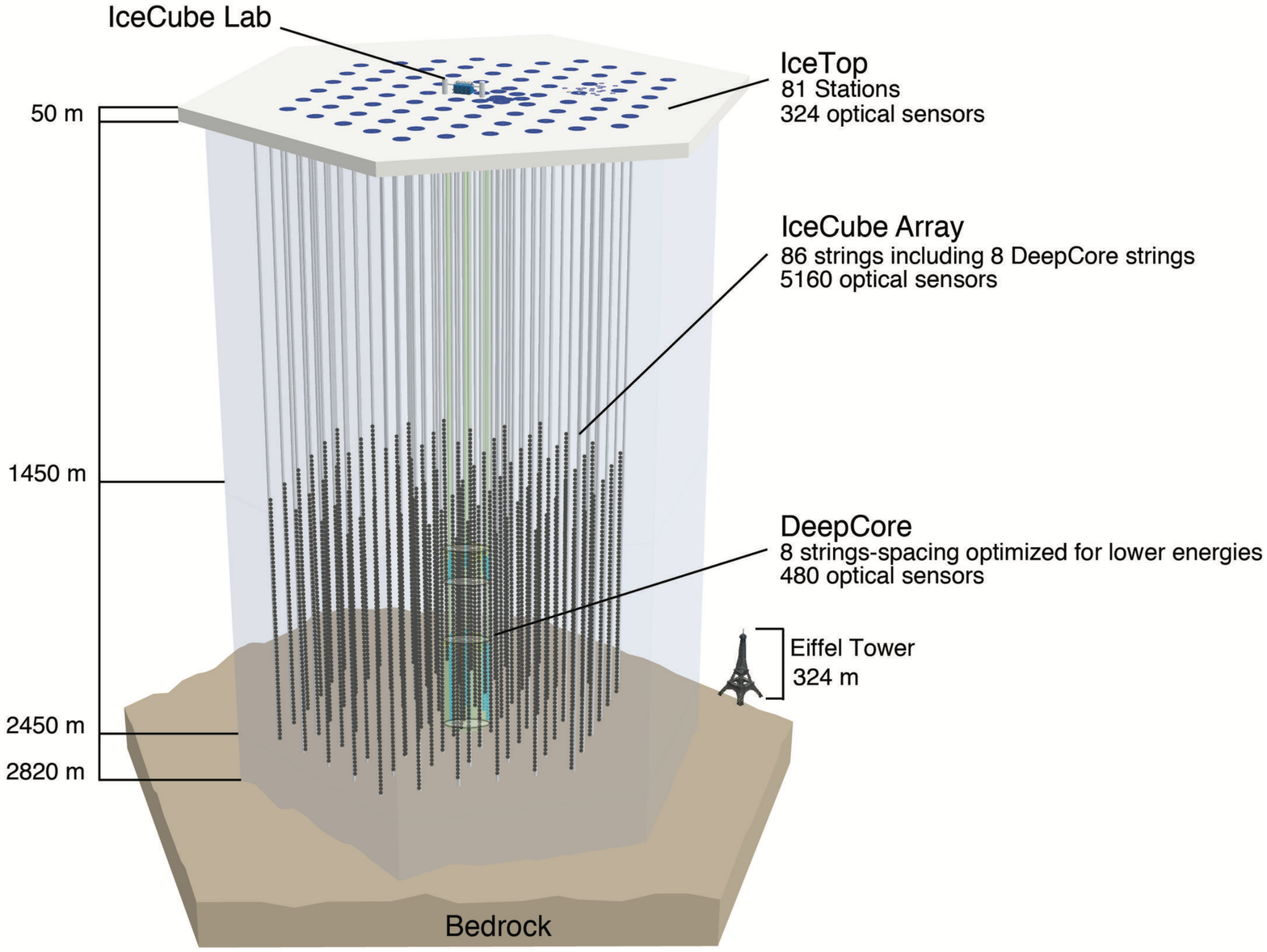}
\caption{Sketch of the IceCube Neutrino Observatory in its final detector configuration. A more compact group of DOMs, named DeepCore, is located at the bottom of the detector. Also shown is IceTop, a surface cosmic ray detector, which consists of 162 water Cherenkov tanks with 324 optical sensors. IceTop data is not used in this work.}
\label{icecube_sketch}
\end{figure}

During May 2010 and May 2011 IceCube detected high-energy particles with 79 strings (IC79) before it operated in its final detector configuration with 86 strings (IC86). 
In this paper, data was taken from the IC79 and four years IC86 (I--IV) configuration. 
A sketch of the final detector configuration can be seen in Figure \ref{icecube_sketch}.

\subsection{Data Sample}
Down-going muons, with an energy of roughly more than 400 GeV, dominate the total trigger rate of 2100 s$^{-1}$ \citep{icecube2017}. 
Due to limited data transfer bandwidth of 100 GB per day from the South Pole to the Northern Hemisphere, online filters are used to reduce the amount of data. 
This paper uses filtered data streams specifically designed to study the Moon and Sun shadows. 
These filters are angular windows ($\pm 10^{\circ}$ in zenith, $\pm180^{\circ}$ in azimuth) around the expected position of Moon and Sun, and are enabled when at least 12 DOMs in three different strings detect photons. 
Furthermore, the Moon and Sun have to reach at least an elevation of 15$^{\circ}$ above the horizon to satisfy the online filters. 
Quality cuts improving the directional reconstruction are applied to all events passing these filters, (see section \ref{sec:qual}). 
The Sun reaches a maximum elevation of $\sim 23.5^{\circ}$ each year, while the Moon's maximum elevation varies between $\sim 24^{\circ}$  and $\sim 18^{\circ}$ at the geographic South Pole for the fraction of the lunar nodal precession covered by these observations. 
The Moon filter is enabled for several consecutive days each month, whereas the Sun filter collects data for approximately 90 days from November through February each season. 

At the South Pole, a fast likelihood-based track reconstruction \citep{Ahrens:2003fg} is performed for each muon event and its direction is compared to the expected position of the celestial body in angular coordinates. 
The data are transferred to the Northern Hemisphere by satellite if the reconstructed event satisfies the Moon or Sun filter. 
Offline, after the events have been transferred, a more sophisticated reconstruction chain is applied. 
The direction of each event is reconstructed using a method similar to the one described in \cite{Aartsen:2013zka}, only instead of a Single-Photo-Electron (SPE) fit, a Multi-Photo-Electron (MPE) fit is performed which uses timing information from all photons rather than just the first photons to arrive. 
The angular uncertainty sigma of the reconstructed direction is computed by then fitting a paraboloid to the likelihood around its minimum and finding the size of the 1-sigma ellipsoidal contour; see \citep{2006APh....25..220N} for more details.

\subsection{Simulations}
\label{sec:simulation}
Simulations are used to verify experimental data, e.g. the zenith angle distribution, and to optimize the used quality cuts (see section \ref{sec:qual}). 
This paper makes use of CORSIKA simulations with a primary energy range from 600 GeV to $10^{11}$GeV and five mass components (p, He, N, Al, Fe). CORSIKA is a Monte Carlo based code that simulates extensive air showers which are produced when cosmic ray particles interact with Earth's atmosphere \citep{1998cmcc.book.....H}. 
This work uses the high-energy hadronic interaction model Sibyll 2.1\citep{PhysRevD.80.094003} and the Antarctic atmospheric profile according to the MSIS-90-E model. 
The CORSIKA sample is weighted with the Gaisser H3a model from \cite{2012APh....35..801G} and compared to the data of the Moon and Sun shadow. 
The simulation sample that triggers the IceCube detector and satisfies the Moon and Sun filters has a median primary particle energy of 40 TeV. 
For 68$\,\%$ of the events in the sample, the energy of the primary particle is between 11 and 200 TeV. 
The CORSIKA generated events are fed into a full detector simulation. 
The same filtering and processing as for experimental data is used. 
For a more detailed description of the simulations, see \cite{Bos2017}.

\subsection{Quality Cuts}
\label{sec:qual}
Events which are mis-reconstructed or have poor angular resolution would reduce the sensitivity of the analysis described in section \ref{sec:binned}. 
Quality cuts are used to remove these events. 
The analysis makes use of two cut variables: the angular uncertainty $\sigma$ of the reconstructed event and a track reconstruction quality estimator, the reduced log-likelihood ($\mathrm{rlogl}$). 
Both cut variables are typically used in point source and other IceCube analyses \citep{2011ApJ...732...18A}. 
Assuming Poisson statistics, the significance $\tilde{S}$ of the shadowing effect is proportional to the square root of the fraction $\eta$ of events passing the cuts and the resulting median angular resolution $\sigma_{\mathrm{med}}$ after cuts \citep{Aartsen:2013zka}, \citep{ap-2-5-2015}:

\begin{align}
\tilde{S} \propto \frac{\sqrt{\eta}}{\sigma_{med}}\,.
\end{align} 
For both IC79 and IC86 the quality cuts are optimized in order to maximize the significance $\tilde{S}$, and the resulting optimal cut values are $\sigma < 0.71^{\circ}$ and $\mathrm{rlogl}<8.1$ \citep{ap-2-5-2015}. 
In an angular window with a size of [72$\times$16]$^{\circ}$ in right ascension and declination around the position of the Moon and Sun, 15--24 million events for the Moon and 40--48 million events for the Sun pass the quality cuts of the analysis each year. 
The number of events that are used in the Moon shadow analysis varies because of the change of the maximum elevation of the Moon each year. 
At higher elevations more muons reach the IceCube detector. 

\clearpage
\section{Binned Analysis}
\label{sec:binned}
A one-dimensional binned analysis of the Moon and Sun shadows is used to study their depth and to estimate the angular resolution of the event reconstruction. 
The analysis observes a profile view of the Moon and Sun shadow and computes the relative deficit of muon events. 
A two-dimensional binned approach uses maps in right ascension and declination to show the shape of the shadows.

\subsection{1-dimensional approach}
\subsubsection{Description of the method}
\label{sec:1dmethod}
The main goal of the 1-D analysis is to observe the profile view of the Moon and Sun shadow and to estimate the angular resolution of IceCube with respect to the studied sample of high-energy muons. 
To measure the deficit in the cosmic ray flux from the direction of the celestial bodies, on- and off-source windows are compared in bins of radial distance $\Psi$ to the Moon/Sun. 
The on-source windows are centered around the expected positions of the Moon and the Sun and have a size of $\pm 5$ degrees. 
The off-source regions are eight windows of the same size with an offset of $\pm 5^{\circ}$, $\pm 10^{\circ}$, $\pm 15^{\circ}$ and $\pm 20^{\circ}$ from the expected positions of the Moon and the Sun in right ascension. 
Because the number of events increases with declination, it is important that the off-source regions have an offset only in right ascension. 
Then, the number of events in each bin in the on-source region is compared with the average number of events in each bin in the eight off-source regions. 

The relative deficit between the number of events in the i-th bin in the on-source region and the average number of events in the same bin in the eight off-source regions is described by:

\begin{align}
      \frac{\Delta N_i}{\langle N \rangle_i} &= \frac{N^{\text{on}}_i - \langle N^{\text{off}}_i \rangle}{\langle N^{\text{off}}_i  \rangle} ,
      \label{eq:on_off}
\end{align}
with  $N^{\text{on/off}}$ being the number of events in on-source/off-source regions. The uncertainty is given by

\begin{align}
      \sigma_{\Delta N_i / \langle N \rangle_i} &= \frac{N^{\text{on}}_i}{\langle N^{\text{off}}_i \rangle} \sqrt{\frac{1}{N^{\text{on}}_i} + \frac{1}{s \cdot \langle N^{\text{off}}_i \rangle}}.
\end{align}  
Here, $s=8$ is the number of off-source regions that are compared with the on-source region. 
In \cite{Aartsen:2013zka} and \cite{ap-2-5-2015} the Moon is expected to be a point-like cosmic ray sink which reduces the muon sample by $\Phi\pi R^2_{\mathrm{M}}$ events. 
Here, $\Phi$ is the cosmic ray flux at the location of the Moon. 
The same approach is used in this paper. 
Here, the radii of the Moon and Sun are described as $R_{\mathrm{M/S}}\approx 0.26^{\circ}$. 
The exact apparent size of the Moon and Sun depends on their distance from Earth and varies between $0.24^{\circ}$ and $0.28^{\circ}$. 
The relative deficit is smeared by the point spread function (PSF) of the IceCube detector.
Under the assumption that the PSF is approximately described by an azimuthally symmetric Gaussian function, it only depends on the radial distance $\Psi$ and is given by \cite{Aartsen:2013zka} as

\begin{align}
	f( \Psi) = - \frac{\Phi\pi {R}^2_{\mathrm{M/S}}}{\sigma_{\mu}^2} \cdot e^{- \Psi^2/2 \sigma_{\mu}^2} .
	\label{eq:psf}
\end{align}
Here, $\sigma_{\mu}$ is the angular resolution of the studied muon sample. 
The relative deficit in the $i$-th bin is:

\begin{align}
	\frac{\Delta N_i}{\langle N \rangle_i}  (\Psi_i)= - \frac{{R}^2_{\mathrm{M/S}}}{2 \sigma_{\mu}^2} e^{- \Psi^2_i / 2 \sigma_{\mu}^2}.
	\label{eq:deficit}
\end{align}
Equation \eqref{eq:deficit} is fitted to IceCube's experimental data. 
The amplitude 
\begin{align}
A = \frac{{R}^2_{\mathrm{M/S}}}{2\sigma_{\mu}^2}
\label{eq:amplitude}
\end{align}
and the angular resolution $\sigma_{\mu}$ are free parameters of the fit. 

We studied the influence of the above fitting procedure on a disk-like sink of cosmic-rays (hence, atmospheric muons) that is smeared by a more realistic point spread function (PSF) of the detector as obtained from the simulations described in section 2.3.
That PSF deviates considerably from the Gaussian approximation made above, especially in the region of large angular errors. 
As a result, inferring the radius ${R}_{\mathrm{M/S}}$ from the fit parameters $A$ and $\sigma_{\mu}$ using equation \eqref{eq:amplitude} induces a bias towards smaller radii, and equation \eqref{eq:amplitude} cannot be used to infer the geometrical radii of the Moon or Sun.

Moreover, as in \cite{Aartsen:2013zka}, the angular size of the Moon and Sun is ignored, which can affect the results of the fit. 
Due to the angular resolution of the IceCube detector, which is on the order of $0.5^{\circ}$, the influence of the angular size of the Moon and Sun should be a few percent at most \citep{Aartsen:2013zka}. 

By comparing the $\chi^2$ of the data points with respect to a constant line $\chi^2_{\mathrm{const}}$, which illustrates the relative deficit without considering any shadowing effects (which is equivalent to assuming a flat background), and the $\chi^2_{\mathrm{Gaussian}}$ of the data points with respect to the fitted Gaussian, the statistical significance is computed:

\begin{align}
    \frac{\Delta\chi^2}{\Delta \text{ndof}} = \frac{\chi^2_{\mathrm{Gaussian}} - \chi^2_{\text{const}}}{\Delta \text{ndof}},
\end{align}
where $\Delta \text{ndof}$ is the difference of the number of degrees of freedom between the flat model and the Gaussian model. Calculating the p-value via $\Delta\chi^2$ results in the significance $S$:

\begin{align}
    S = \sqrt{2} \cdot \text{erf}^{-1}(1-p).
\end{align}

\subsubsection{Results}
\label{sec:1dresults}
After the quality cuts are applied to IceCube's experimental data, approximately 30$\,\%$ of all muon events passing Moon and Sun filters survive these cuts each year and are used to analyze the cosmic ray Moon and Sun shadow. 
Results for the 1-D analysis can be found in Table \ref{tab:results_1d}. 
Keep in mind that for the reasons described in section \ref{sec:1dmethod} the fit parameters $A$ and $\sigma_{\mu}$ in Table \ref{tab:results_1d} cannot be used to infer the geometrical radii of Moon and Sun.

The cosmic ray Moon shadow is observed with high statistical significance ($>11\sigma$) in every year. 
The angular resolution of the Moon shadow analysis, given by the width of the fitted Gaussian, remains stable for the entire observation period. 
IceCube's median angular error $\sigma_{\mathrm{med}}$, as derived from simulations, is 0.68$^{\circ}$, and 68$\,\%$ of the events are contained between 0.29 and 1.62 degrees. 
The data analysis shows better angular resolution compared to the median angular error, which is expected and was already observed in \cite{Aartsen:2013zka}.

Furthermore, the amplitude varies only within its statistical uncertainties. 
These results show that the IceCube detector is operating stably for the five-year observation period. 
In \cite{Aartsen:2013zka}, where data are used from two years in which IceCube operated in smaller detector configurations, the median angular resolution was measured to be $(0.71\pm 0.07)^{\circ}$ and $(0.63\pm 0.04)^{\circ}$ for the 40 string configuration in 2008/2009 and the 59 string configuration in 2009/2010, respectively. 
The differences between IC40, IC59 and this analysis are expected, because this work makes use of the final detector configuration with 86 strings and the previous configuration with 79 strings. 

\begin{table}[htbp]
\centering
\begin{tabular}{c|c|c|c|c|c|c|c}
& Year  &  $\sigma_{\mu} [^{\circ}]$ & $A$ & S$[\sigma]$\\
\hline\hline
&2010/11 			&			0.43	$\pm$	0.05	&	0.12	$\pm$	0.02 & $>12$ \\		
&2011/12			&			0.45	$\pm$	0.05	&	0.13	$\pm$	0.02	& $>13$  \\
Moon&2012/13	&			0.49 $\pm$	0.05 &	0.11 $\pm$	0.02 & $>12$ \\
&2013/14			&			0.47 $\pm$ 0.05 &	0.12 $\pm$	0.02 & $>12$ \\
&2014/15			&			0.47 $\pm$	0.06 &	0.13 $\pm$	0.02 & $>11$ \\
\hline\hline
&2010/11 				&			0.53 $\pm$	0.05 &	0.11	$\pm$	0.01 & $>16$ \\
&2011/12 				&			0.49	$\pm$	0.06	&	0.08	$\pm$	0.01 & $>13$ \\
Sun&2012/13 			&			0.57	$\pm$	0.05	&	0.09 $\pm$	0.01 & $>16$ \\
&2013/14 				&			0.58	$\pm$	0.07	&	0.05	$\pm$	0.01 & $>10$ \\
&2014/15 				&			0.57	$\pm$	0.07	&	0.06	$\pm$	0.01 & $>10$ \\
\end{tabular}
\caption{Results of the 1-Dimensional analysis, with A as the amplitude and $\sigma_{\mu}$ as the width of the Gaussian.}
\label{tab:results_1d}
\end{table}

\begin{figure*}[htbp]
\centering
\begin{tabular}{cc} 
\includegraphics[trim = 5mm 0mm 5mm 0mm, clip, width=0.45\textwidth]{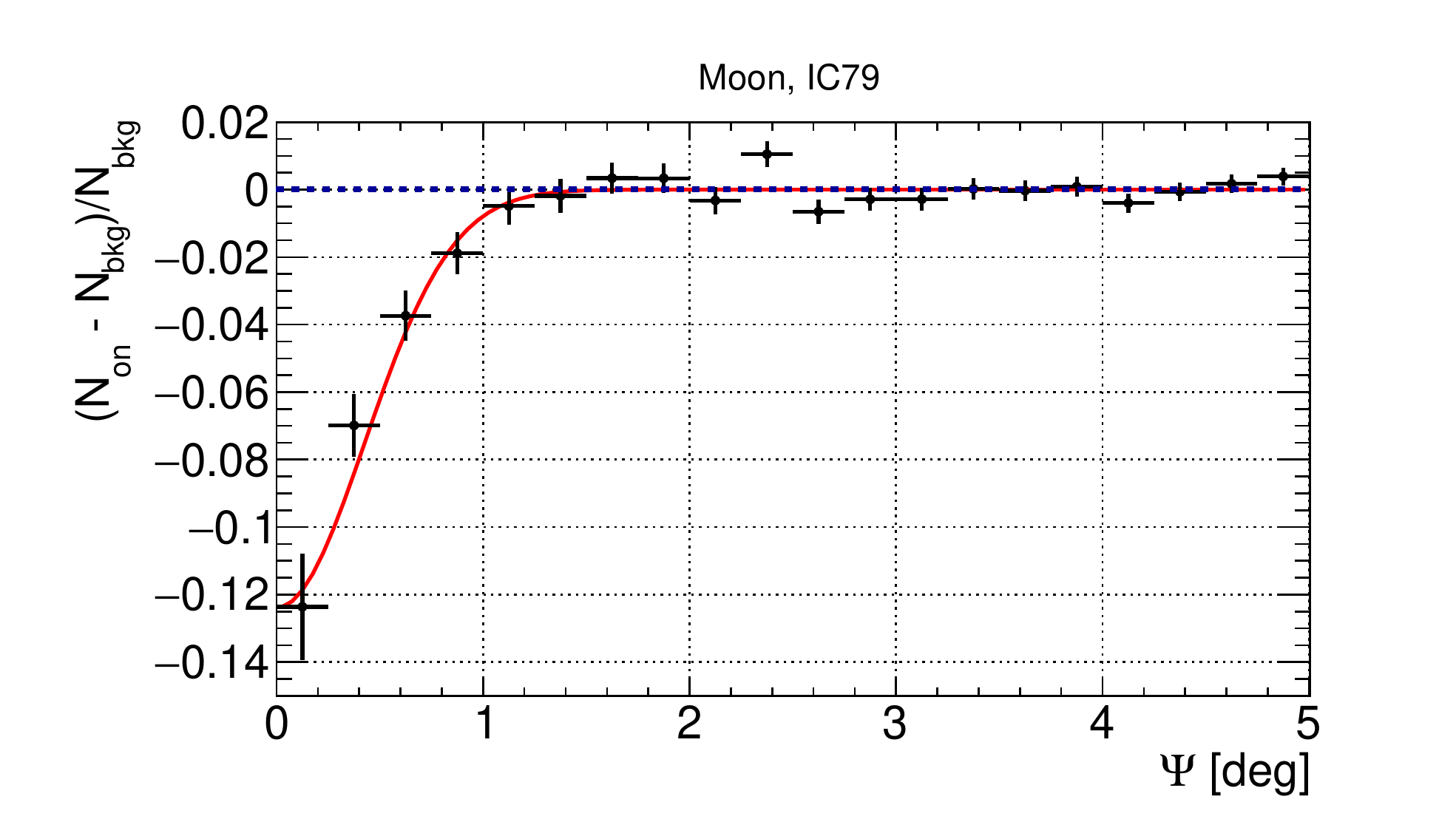}  & \includegraphics[trim = 5mm 0mm 5mm 0mm, clip, width=0.45\textwidth]{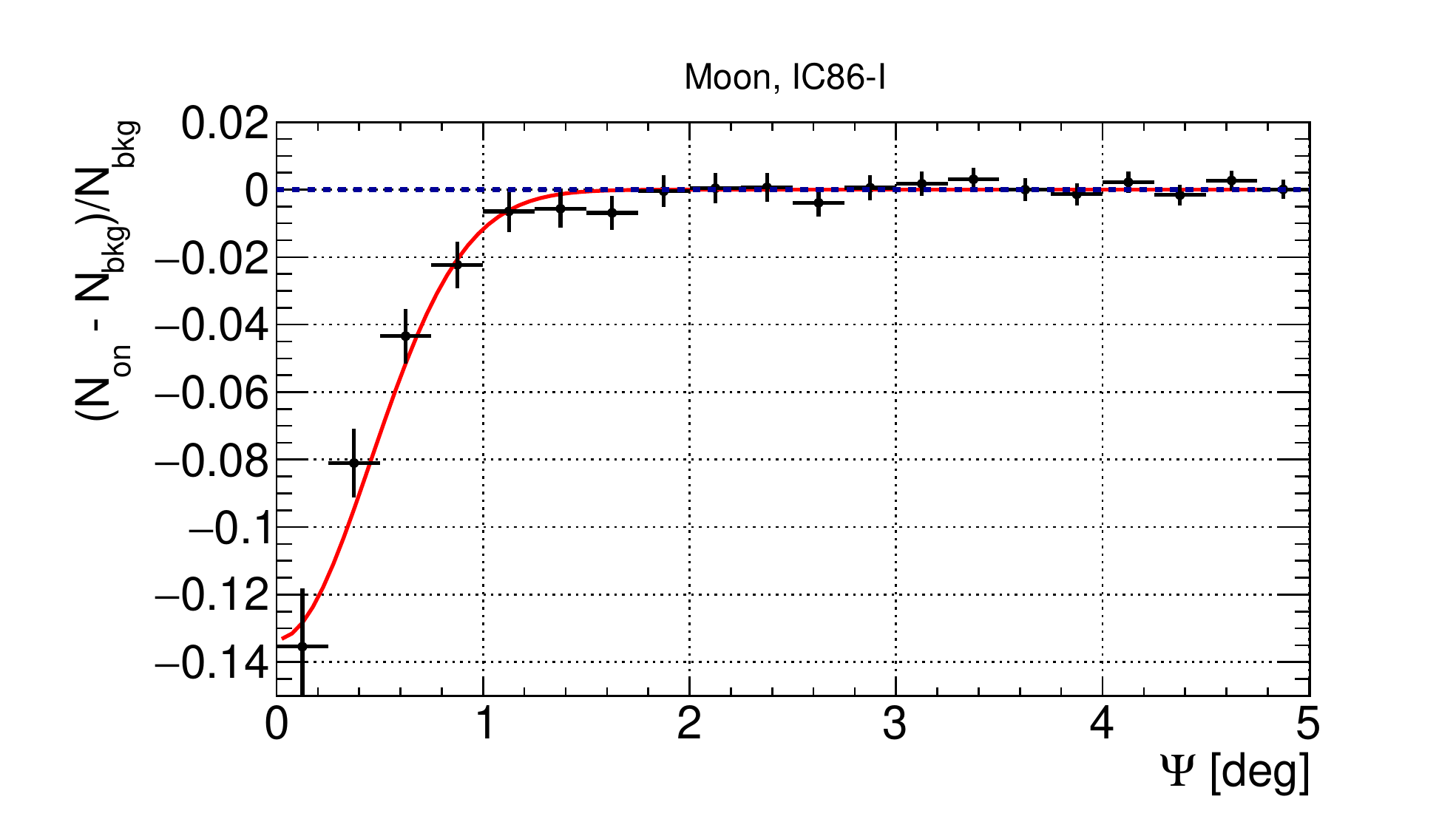} \\ 
\includegraphics[trim = 5mm 0mm 5mm 0mm, clip, width=0.45\textwidth]{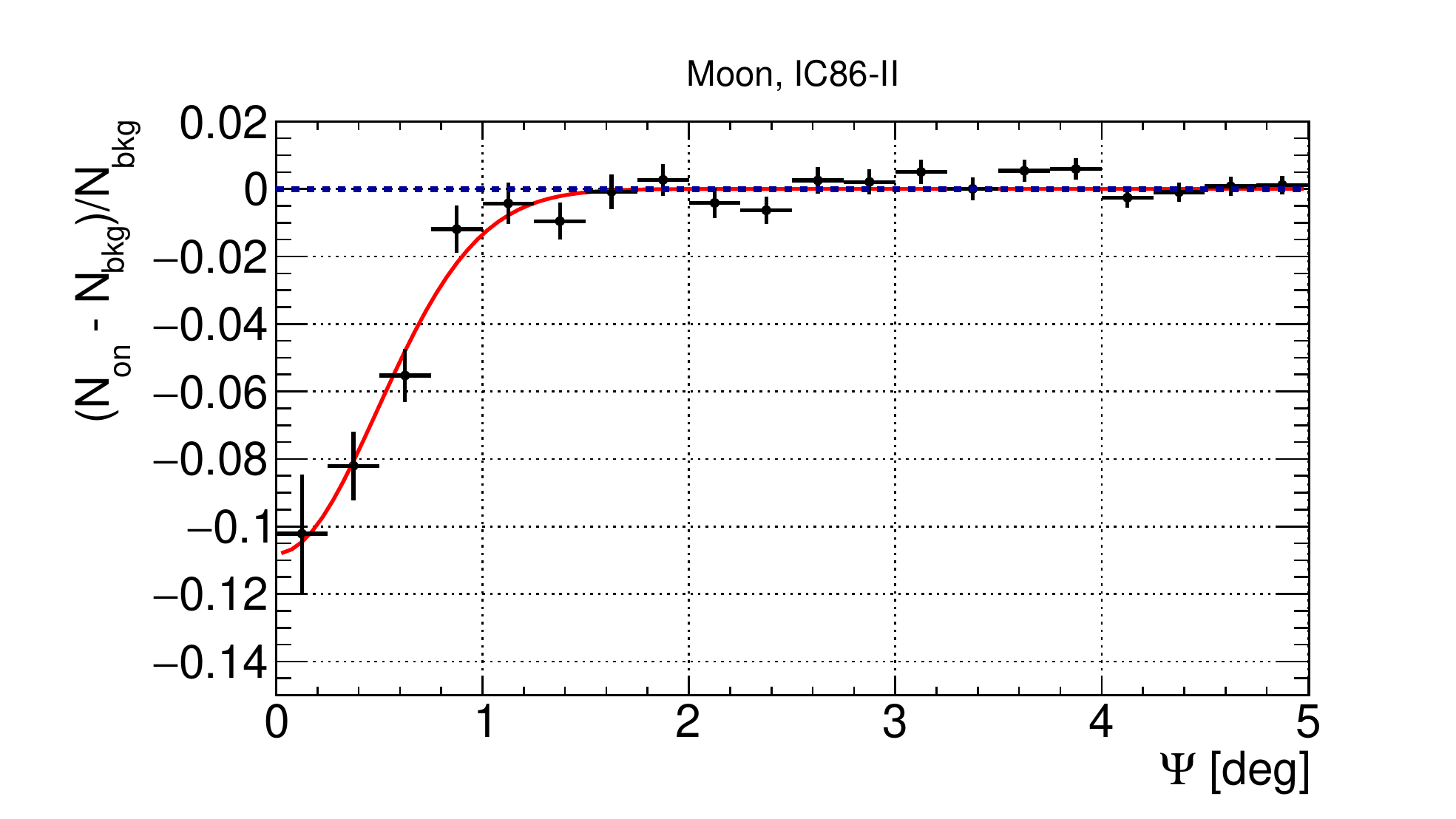} & \includegraphics[trim = 5mm 0mm 5mm 0mm, clip, width=0.45\textwidth]{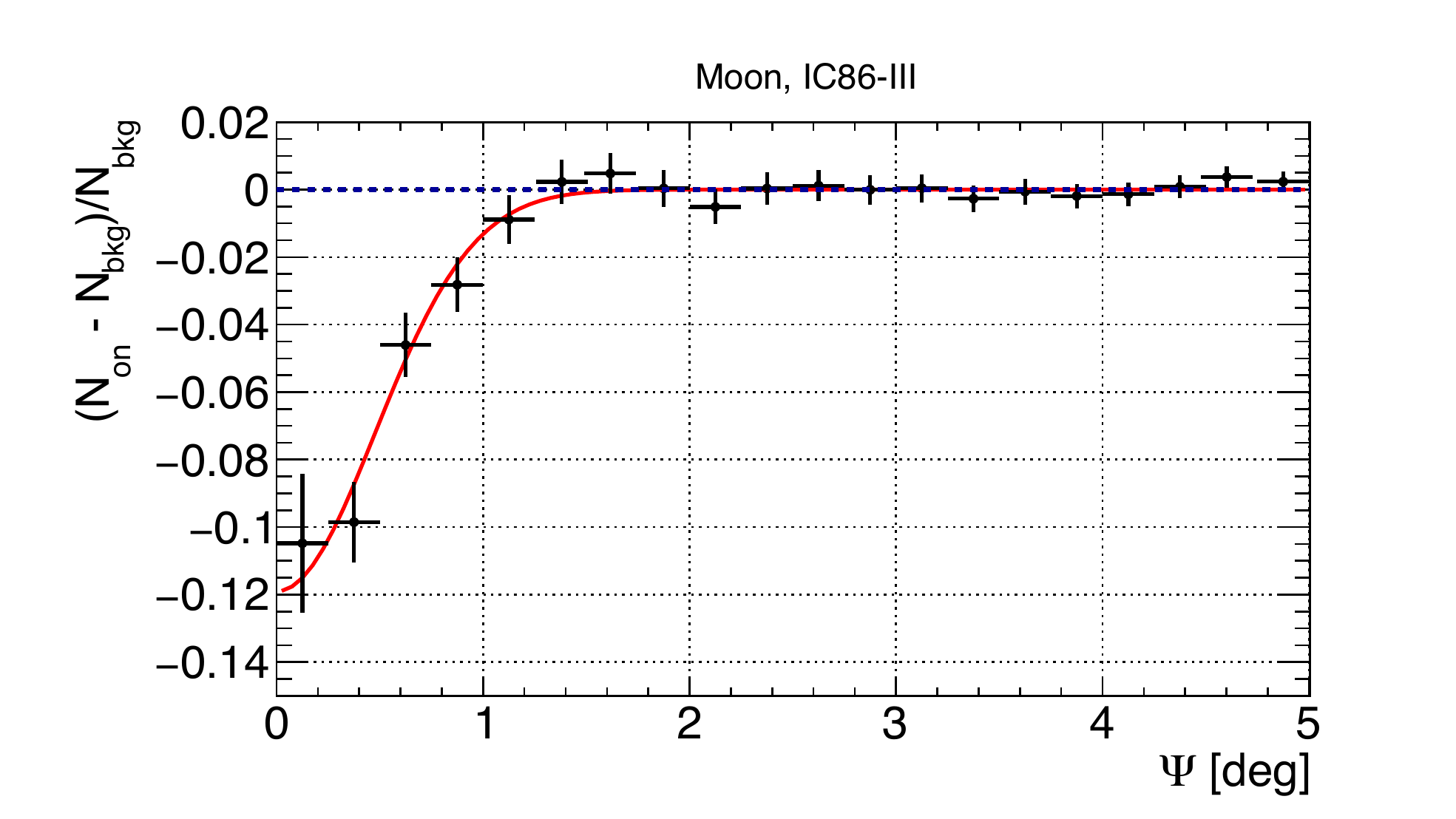}
\end{tabular}
\centering 
\includegraphics[trim = 5mm 0mm 5mm 0mm, clip, width=0.45\textwidth]{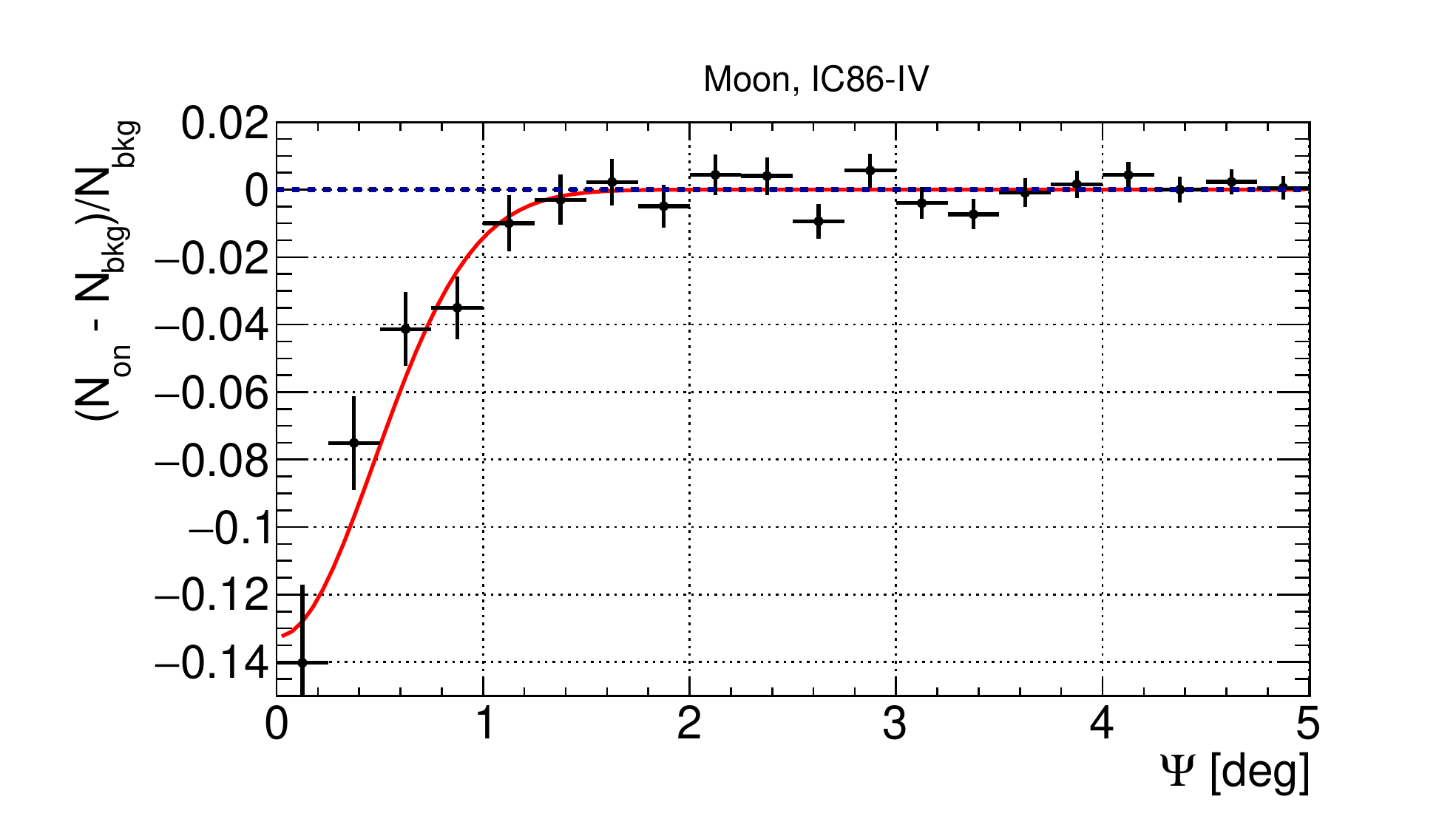} 
\caption{Observed cosmic ray deficit due to the Moon shadow inferred with the binned analysis in one dimension. The relative deficit in the number of events in on-source and off-source regions of each bin is shown for each year as a function of the radial distance $\Psi$ to the Moon. The amplitude A and the width $\sigma_{\mu}$ remain constant, which verifies that the IceCube detector operates stably.}
\label{fig:1d_moon}
\end{figure*}

In Figure \ref{fig:1d_moon} the 1-D analysis results for the Moon shadow are shown for the five-year observation period. 
The y-axis illustrates the relative deficit of events in each bin of the on-source and off-source regions as a function of the angular distance on the x-axis. 
A Gaussian (red line) with parameters from Table \ref{tab:results_1d} and a line for no relative deficit (blue line) are also shown. 

\begin{figure*}[htbp]
\centering
\begin{tabular}{cc} 
\includegraphics[trim = 5mm 0mm 5mm 0mm, clip, width=0.45\textwidth]{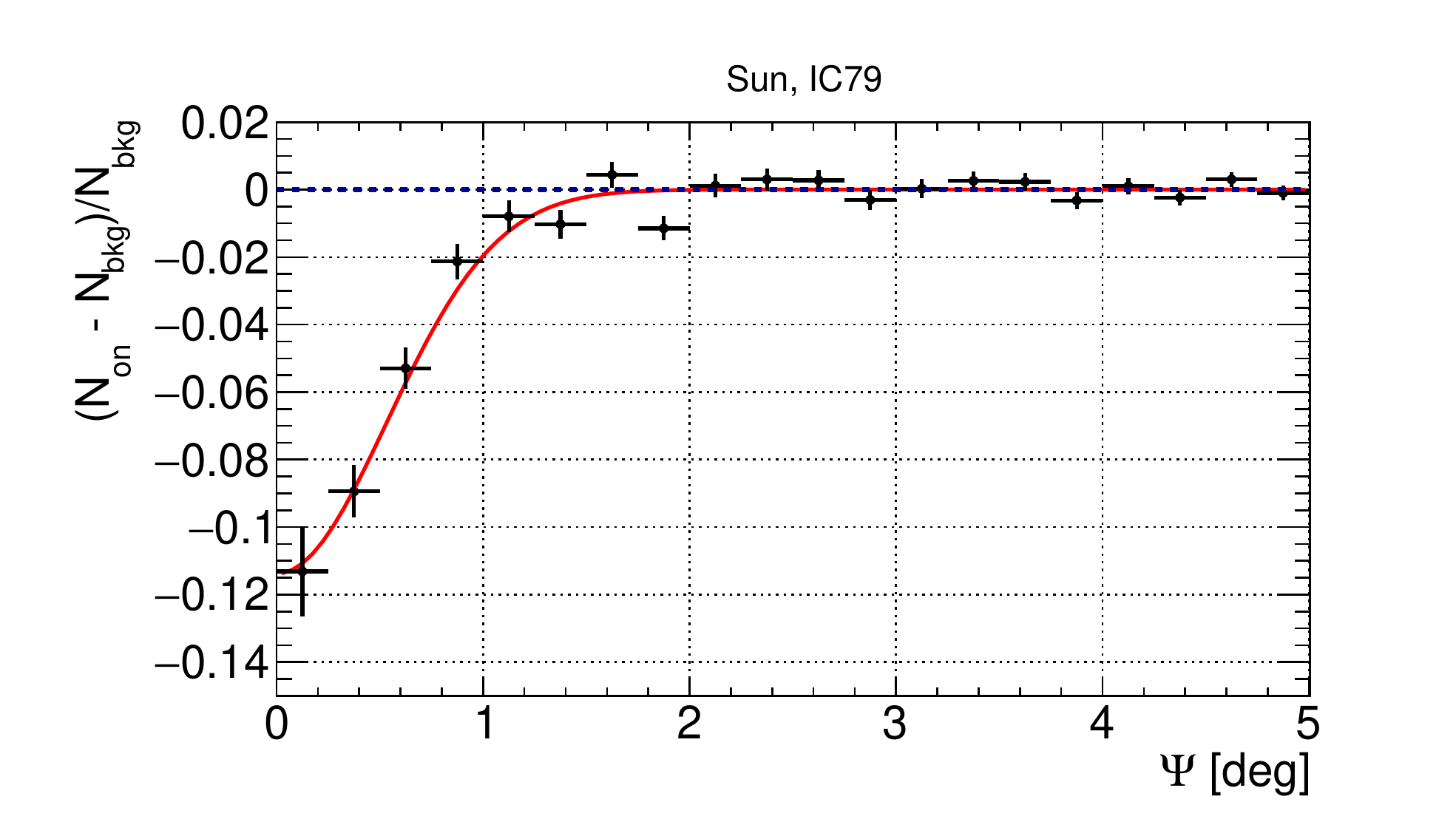}  & \includegraphics[trim = 5mm 0mm 5mm 0mm, clip, width=0.45\textwidth]{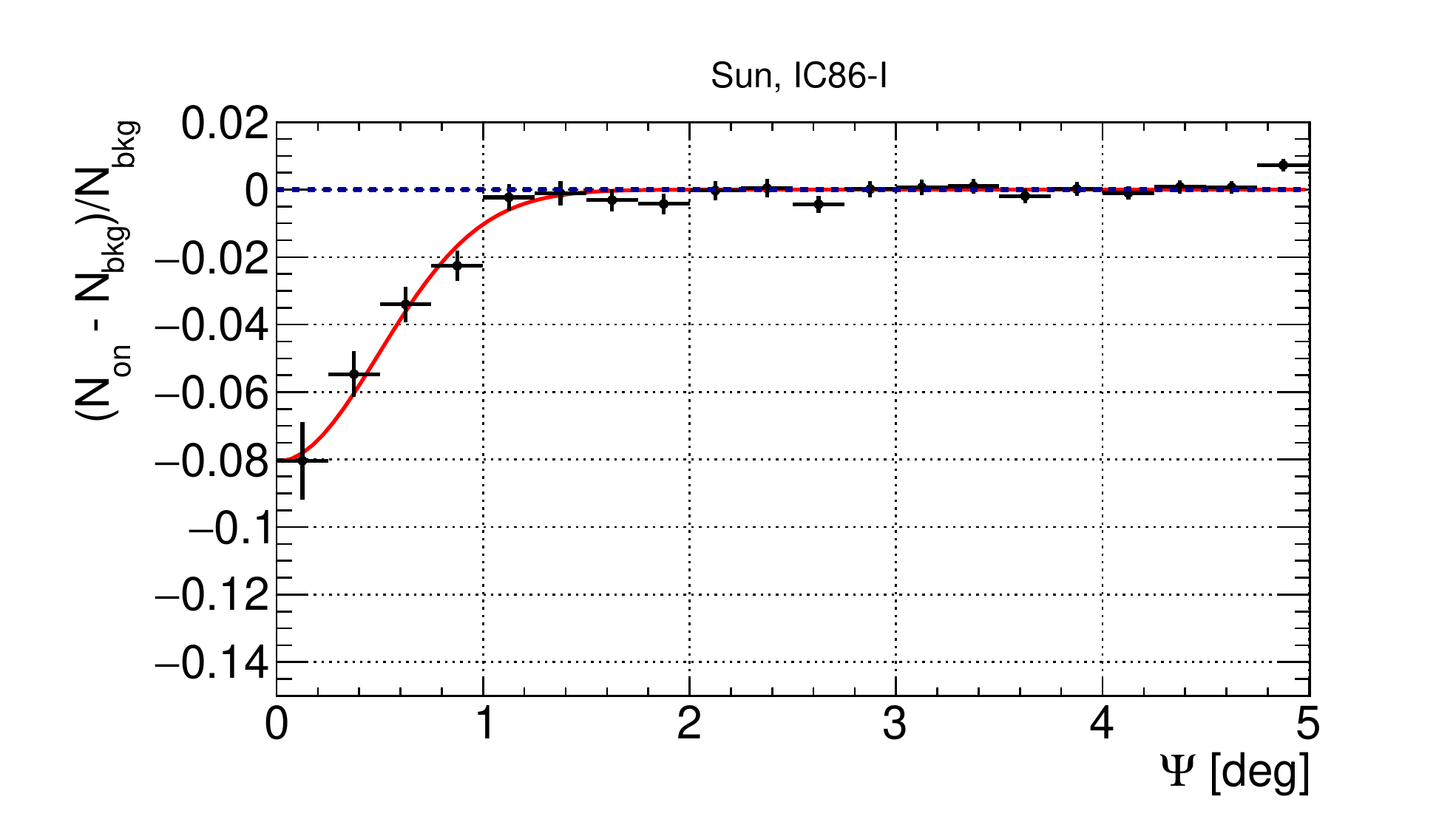} \\
\includegraphics[trim = 5mm 0mm 5mm 0mm, clip, width=0.45\textwidth]{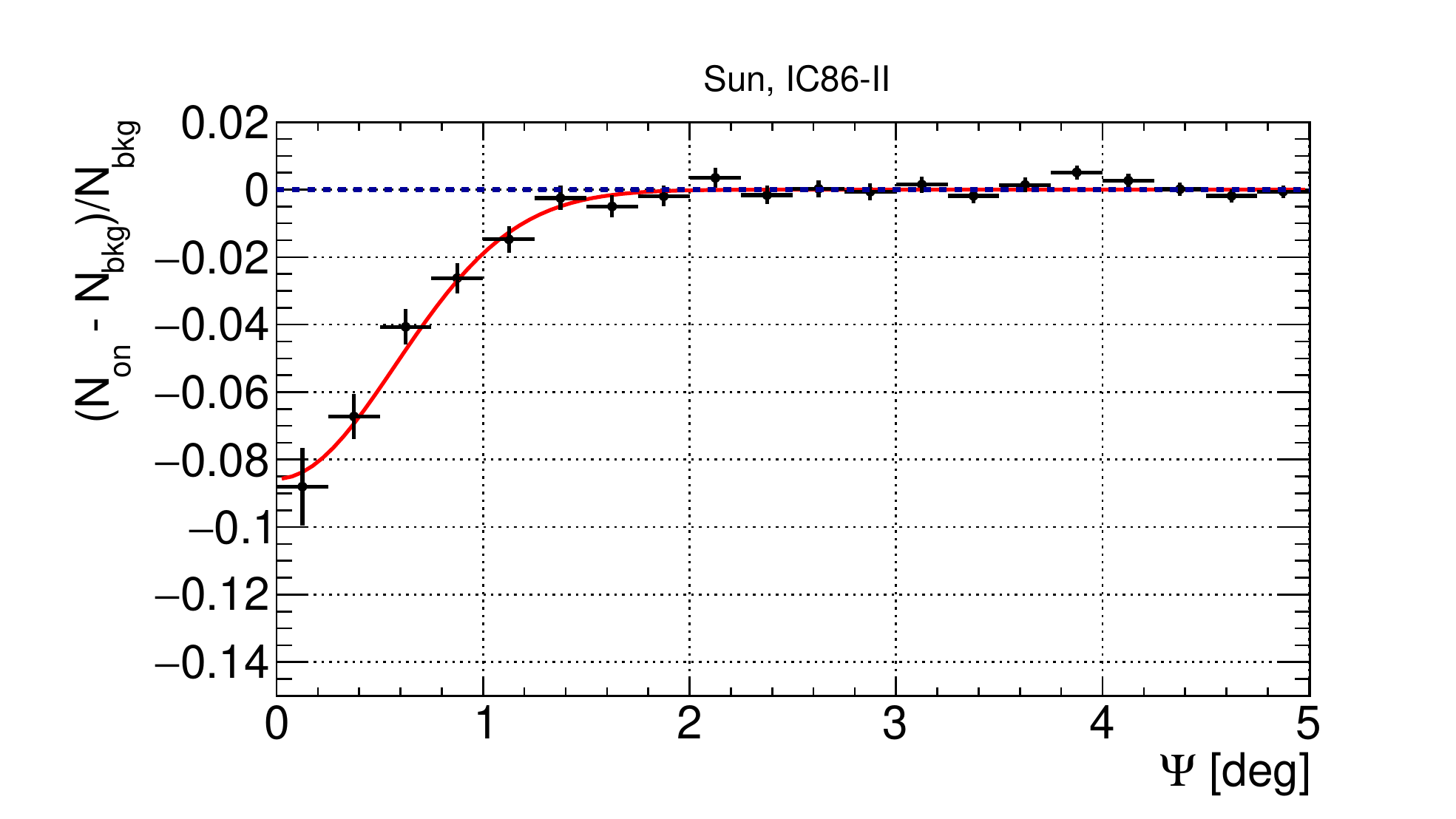} & \includegraphics[trim = 5mm 0mm 5mm 0mm, clip, width=0.45\textwidth]{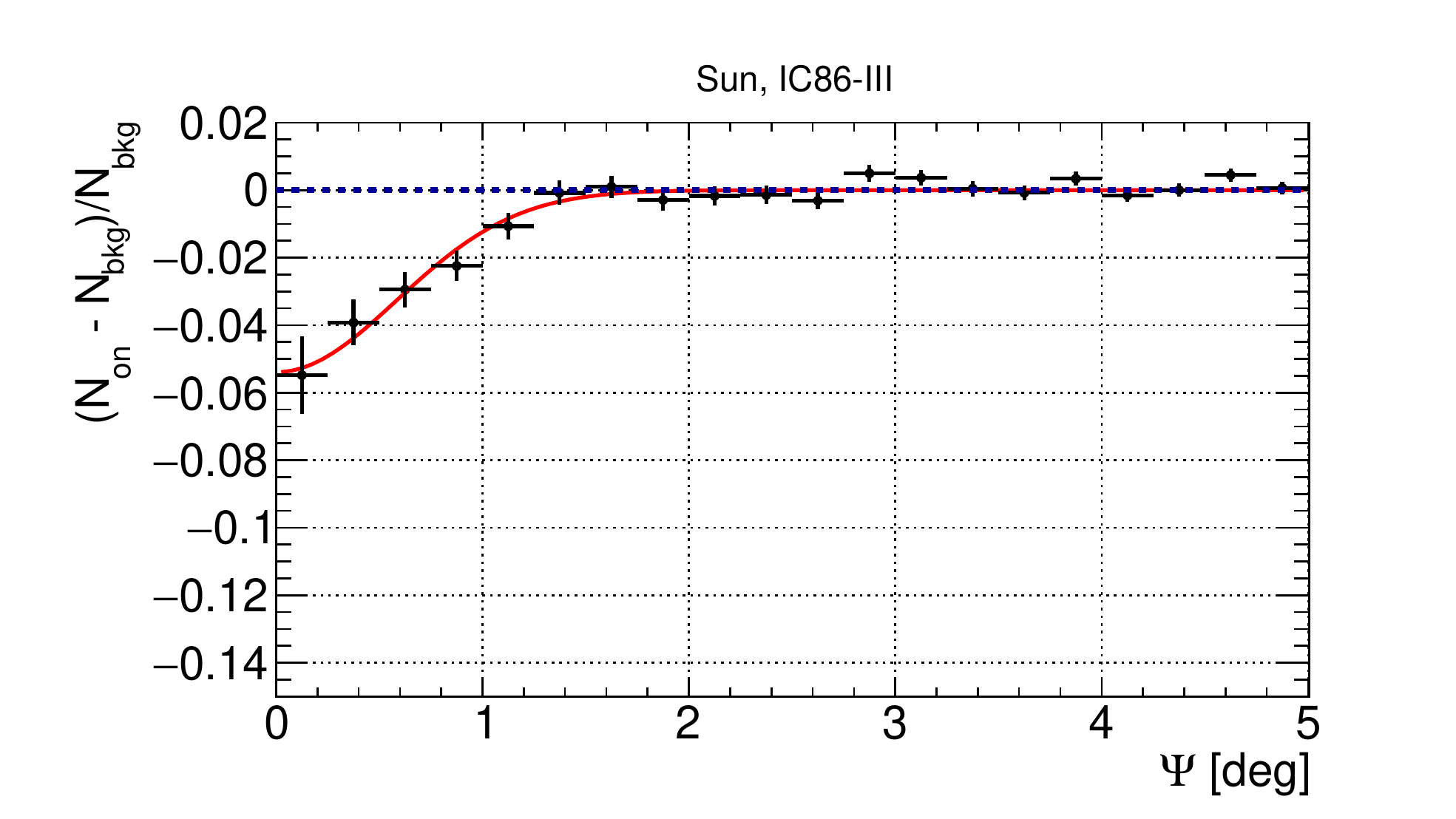}
\end{tabular}
\centering
\includegraphics[trim = 5mm 0mm 5mm 0mm, clip, width=0.45\textwidth]{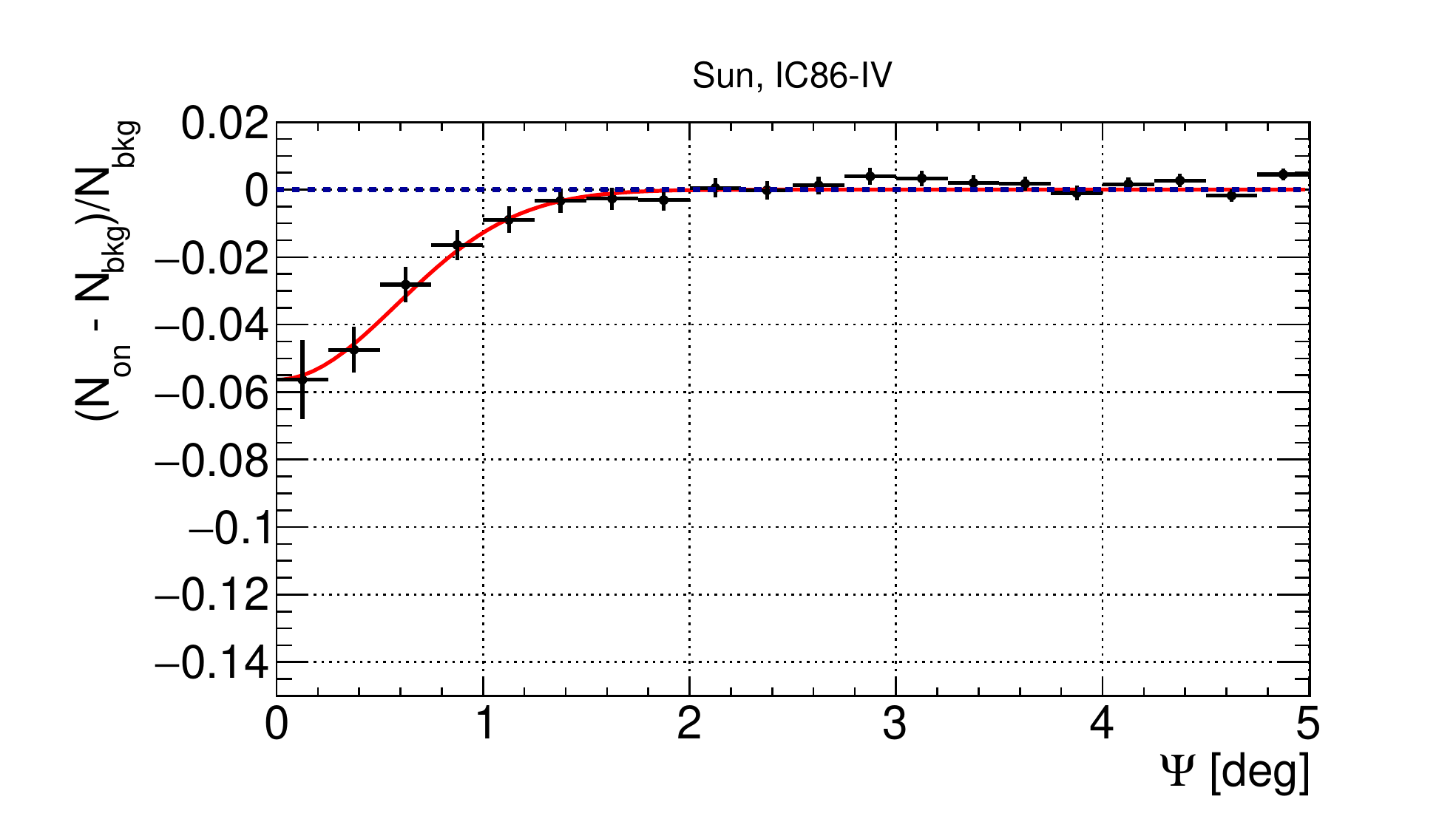} 
\caption{Observed cosmic ray deficit due to the Sun shadow inferred with the binned analysis in one dimension. The relative deficit in the number of events in on-source and off-source regions of each bin is shown for each year as a function of the radial distance $\Psi$ to the Sun. The amplitude varies during the five-year observation time.}
\label{fig:1d_sun}
\end{figure*}

The 1-D analysis is performed for the Sun shadow analysis for the same observation period (see Table \ref{tab:results_1d}). 
As we derive quantitatively from the width and amplitude of the shadow (see Table \ref{tab:results_1d}), the angular resolution is consistent with a constant value for the five-year observation period. 
The amplitude of the Sun shadow, on the other hand, varies between $(11\pm1)\,\%$ and $(5\pm1)\,\%$.
For the Sun, the interpretation of $A$ and $\sigma_{\mu}$ is more complex than for the Moon.
Besides the detector effects that cause the smearing in the case of the Moon shadow, we also expect physical solar effects to be relevant in the case of the Sun shadow.
For example, the deflection of cosmic rays in the solar magnetic field can presumably change the values measured for $A$ and $\sigma_{\mu}$.
In this paper, we do not aim to separate the effects of absorption and deflection on the Sun shadow.
The time variability of $A$ that can be seen in Table \ref{tab:results_1d}, however, must be intrinsic to the Sun since for the Moon both fit parameters remain stable during the observation period.
Figure \ref{fig:1d_sun} shows the measured one dimensional Sun shadow for each of the five years.   

\subsection{2-dimensional approach}
\subsubsection{Description of the method}
In a second approach, two-dimensional maps of the cosmic ray shadows of the Moon and Sun are created. 
The goal is to make the shadowing effects of the Moon and Sun visible in a three dimensional representation, with the magnitude of the shadowing effect as the third dimension. 
This approach compares each bin of an on-source region with the average of the same bins of two off-source regions with an offset of $\pm 18^{\circ}$ in right ascension. 
Due to a higher muon trigger rate from higher declinations, the off-source regions only have an offset in right ascension. 
A map with a size of $[6^{\circ} \times 6^{\circ}]$ around the expected position of the celestial bodies is used to show relative declination, relative right ascension and relative deficit of events in each bin. 

After assigning muon events to on- and off-source regions, a smoothing method, which takes the average number of events within a chosen rebinning radius of $0.35^{\circ}$ of each bin in the  $[6^{\circ} \times 6^{\circ}]$ window, is applied in order to make the shadowing effects visible. 
In order to maximize the statistical significance, a bigger rebinning radius can be used. 
However, the goal of this approach is to show the shadowing effects of Moon and Sun with the highest relative deficit. 
Thus, the rebinning radius is chosen to be smaller than the angular size of Moon and Sun. 
With a rebinning radius of $0.35^{\circ}$ the lower limit is reached. 
Smaller radii would lead to shadowing effects with insufficient statistical significance. 
Similar methods to analyze the shadowing effects of the Moon and Sun are used in e.g. \cite{hawc2013} and \cite{2013PhRvL.111a1101A}. 

\subsubsection{Results}
\begin{figure*}[htbp]
	\centering
		\begin{tabular}{cc} 
\includegraphics[trim = 5mm 0mm 5mm 0mm, clip, width=0.4\textwidth]{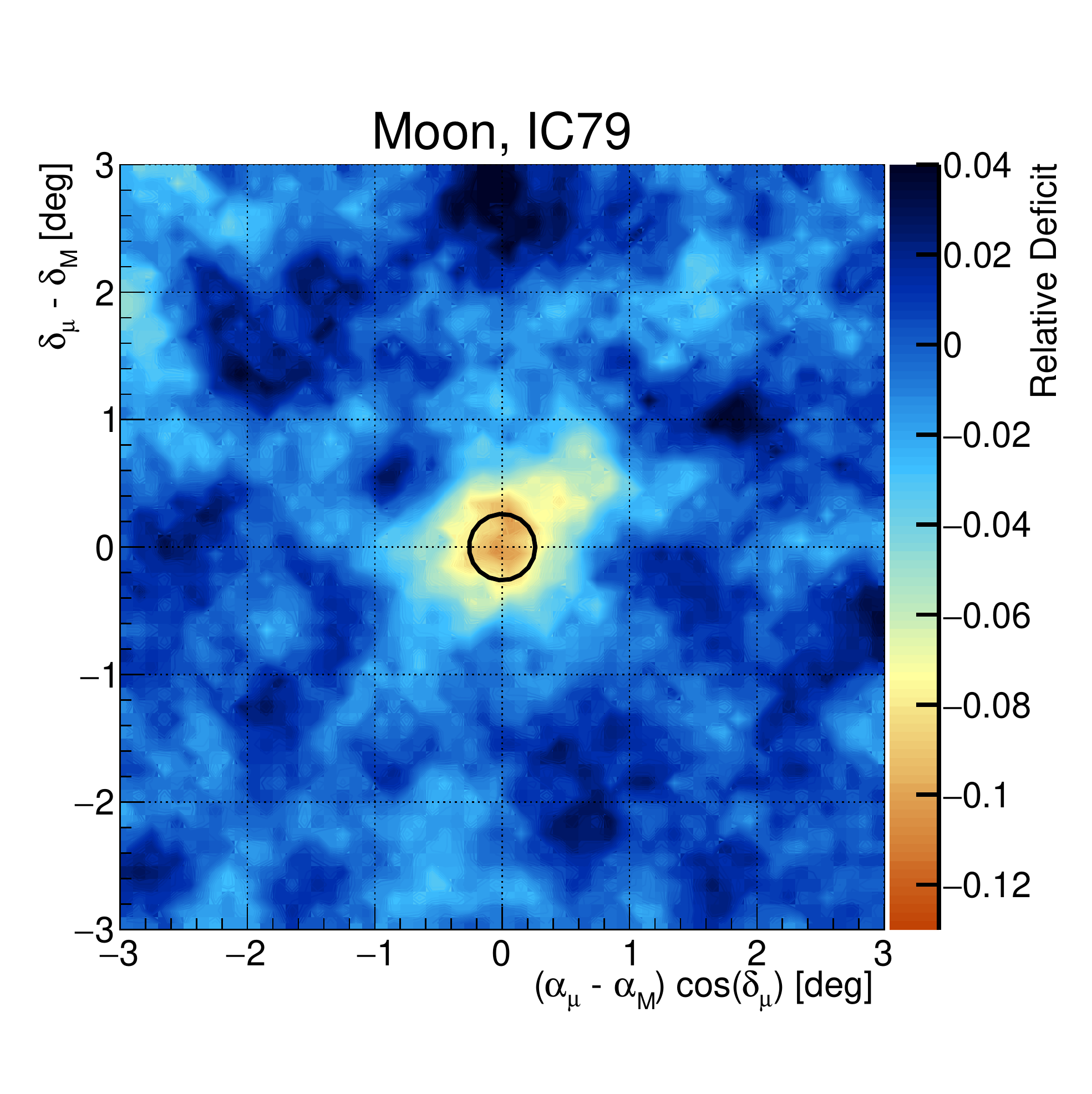}  & \includegraphics[trim = 5mm 0mm 5mm 0mm, clip, width=0.4\textwidth]{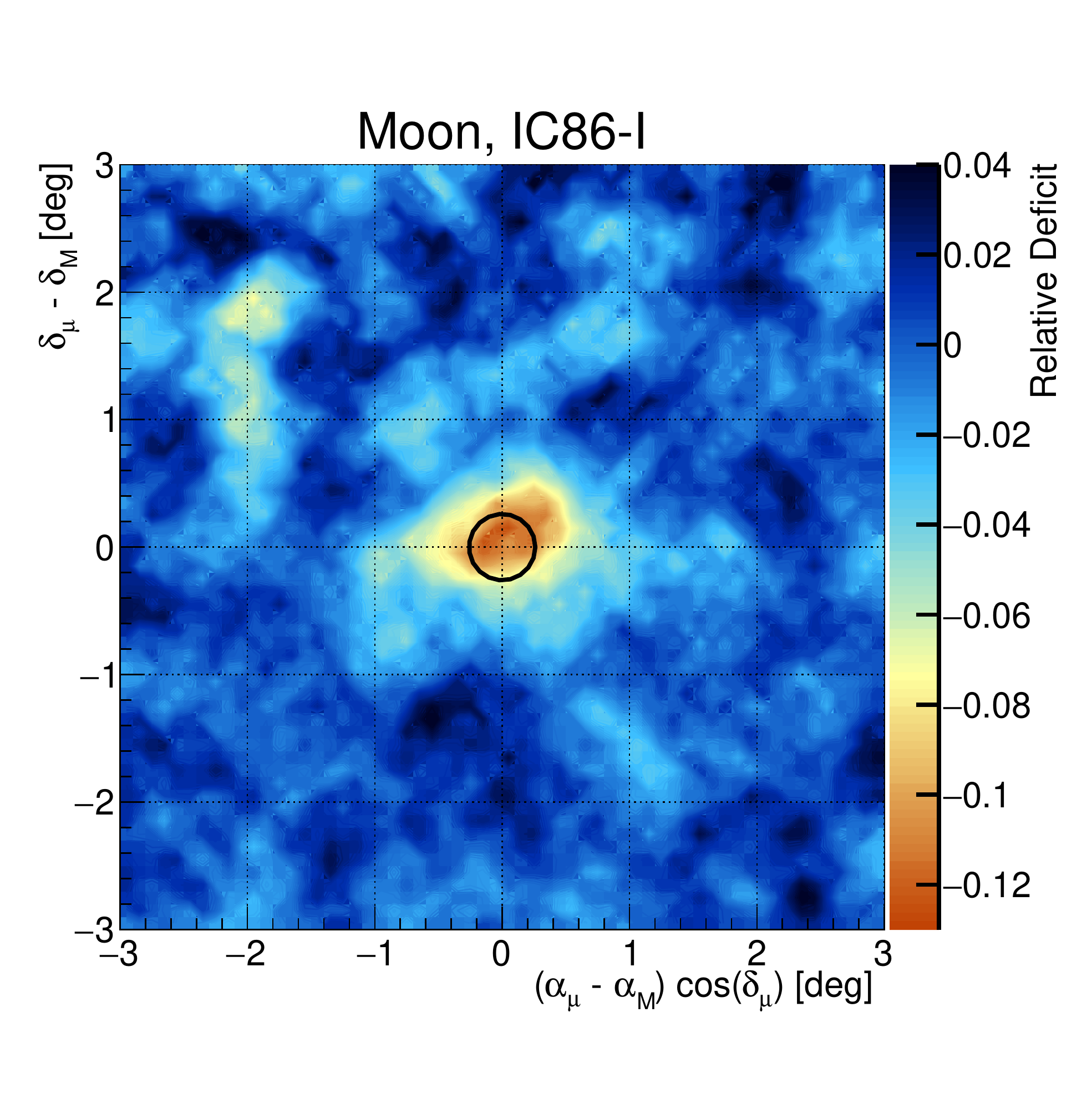} \\
\multicolumn{2}{c}{ \includegraphics[trim = 5mm 0mm 5mm 0mm, clip, width=0.4\textwidth]{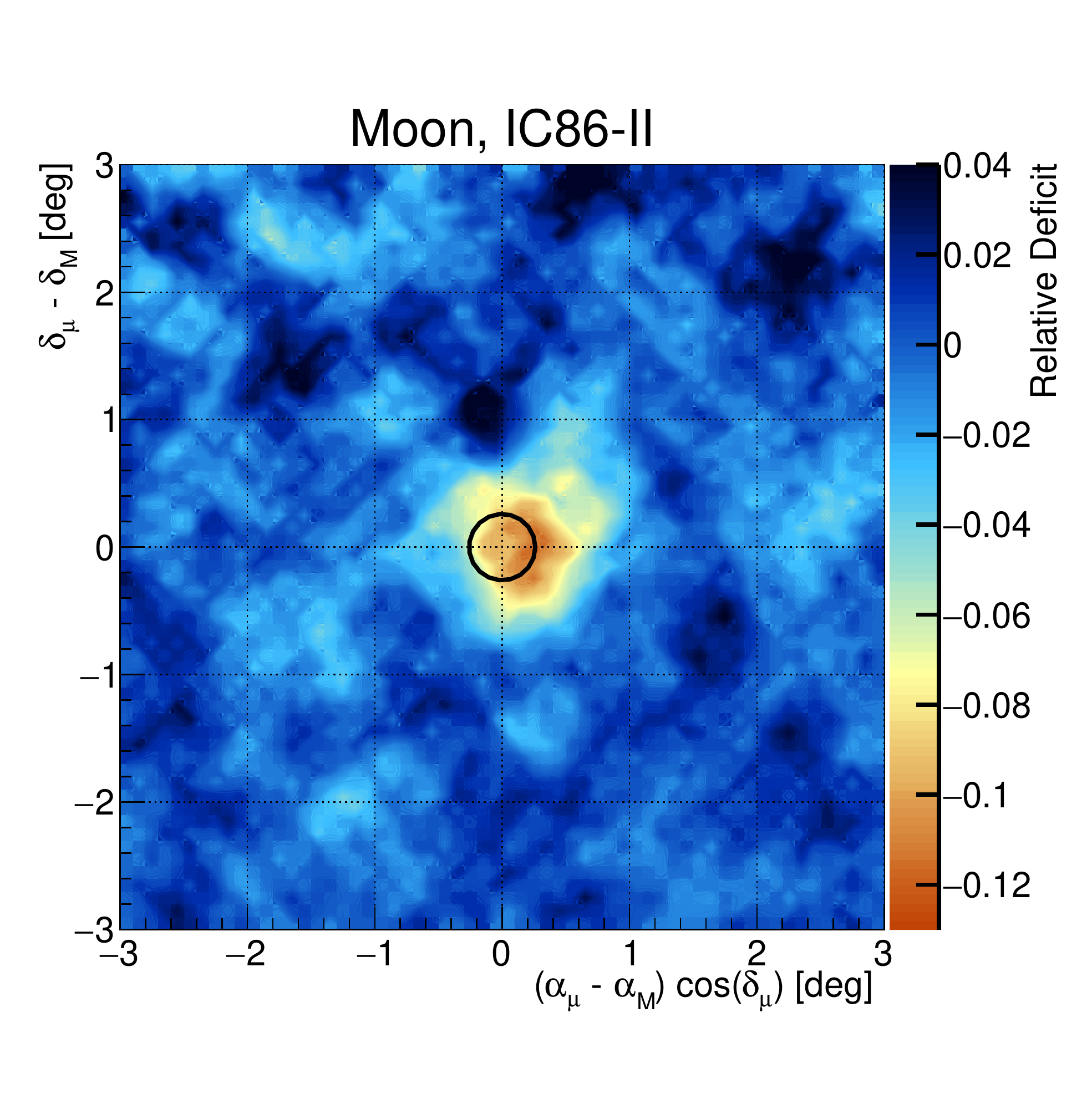} \hspace{0.23cm} \includegraphics[trim = 5mm 0mm 5mm 0mm, clip, width=0.4\textwidth]{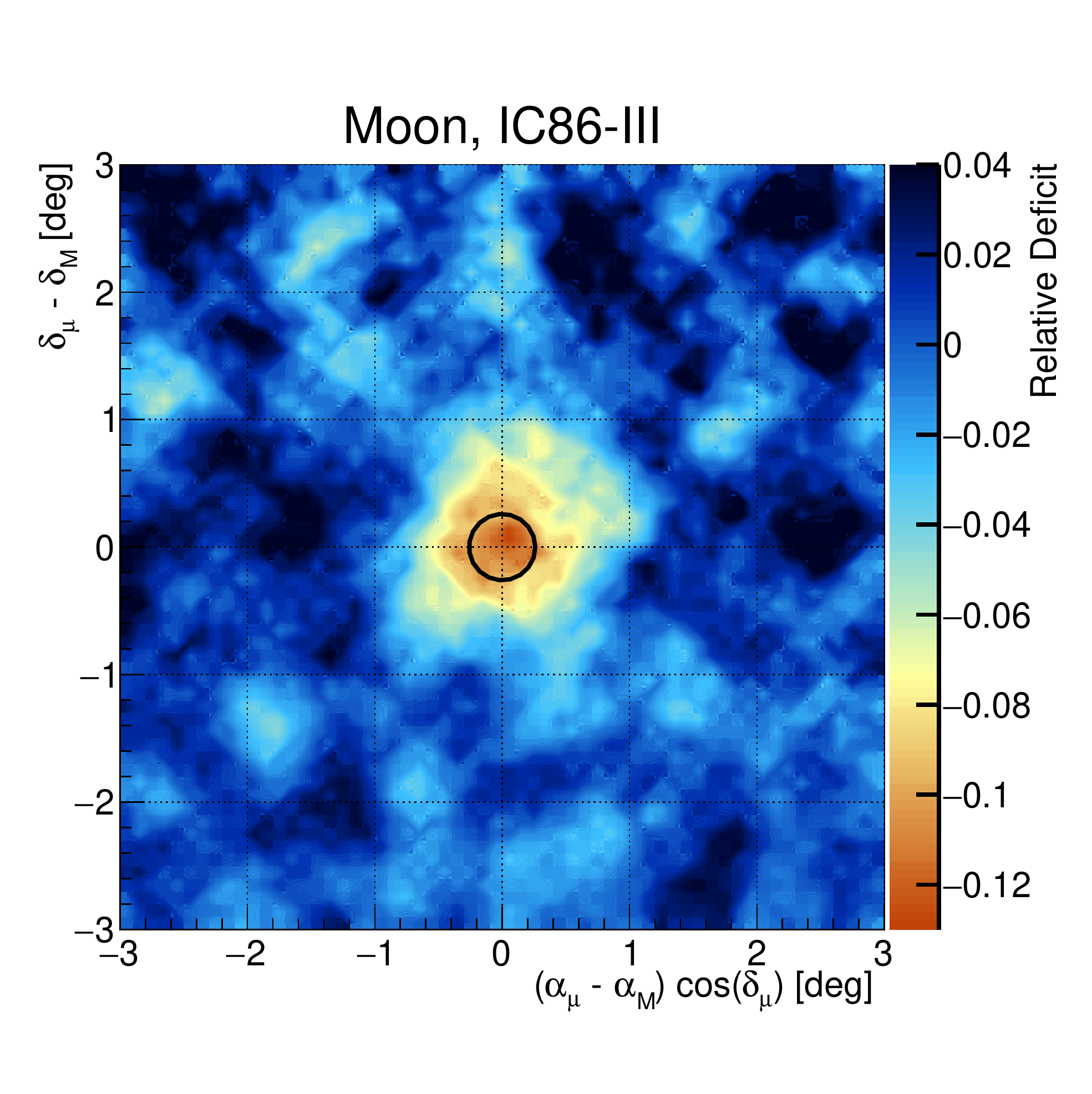}} \\
\multicolumn{2}{c}{\includegraphics[trim = 5mm 0mm 5mm 0mm, clip, width=0.4\textwidth]{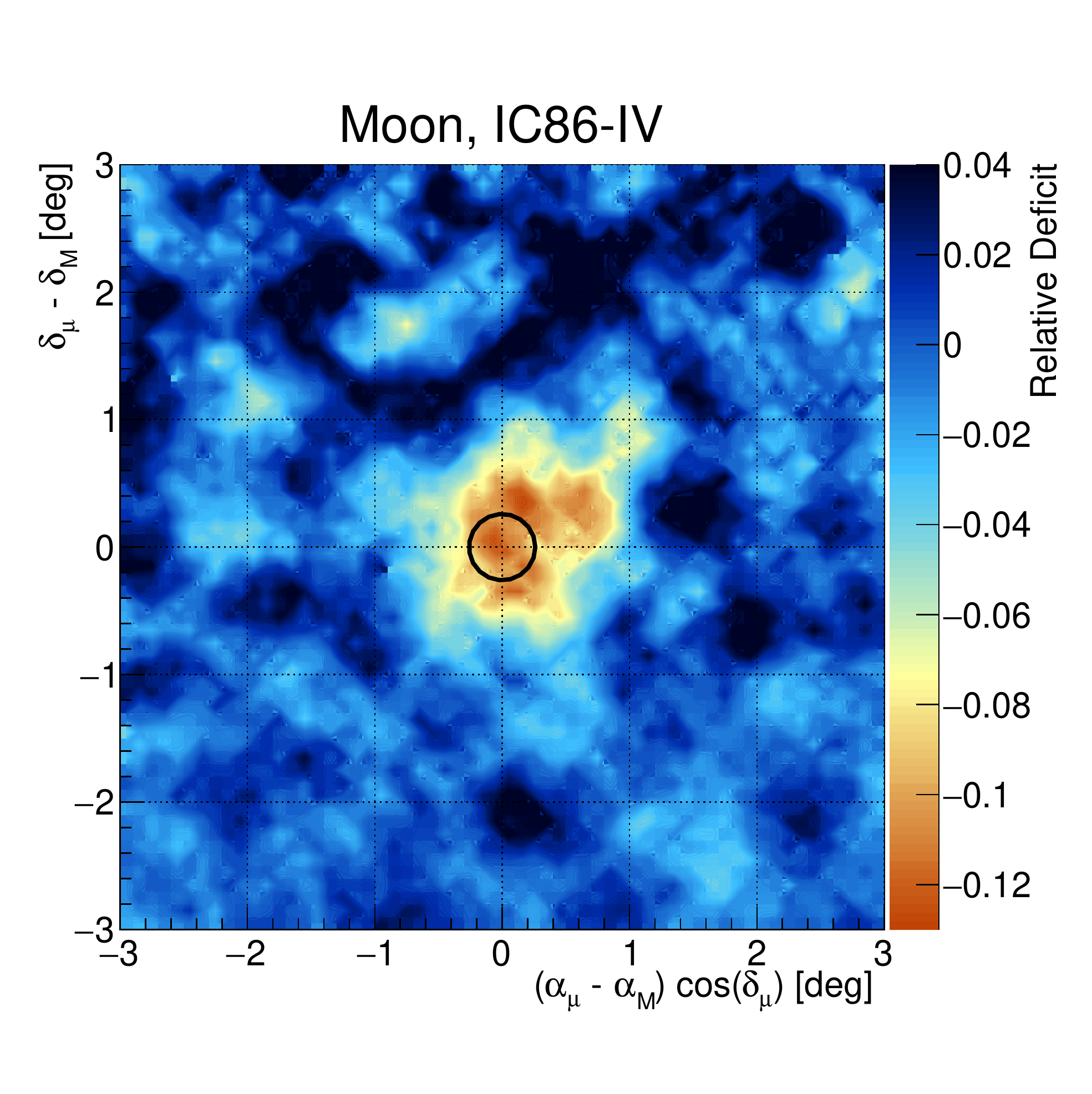}} \\
\end{tabular}
	\caption{Results for the binned visualization of the Moon shadow in two dimensions. The expected position of the Moon is in the middle of the $[6^{\circ} \times 6^{\circ}]$ plot. The black circle shows the actual size of the Moon. A smoothing radius has a size of $0.35^{\circ}$. Each map shows the cosmic ray Moon shadow for a season from 2010 through 2015. The contour represents the relative deficit of events in each bin. The shadowing remains stable for the entire observation period, which is expected for a stably operating detector.}
	\label{fig:2d_moon}
\end{figure*}

Figure \ref{fig:2d_moon} shows the results of the two-dimensional approach for the IC79 and four years of the IC86 (I--IV) configuration. 
The black circle illustrates the actual size of the Moon of $0.5^{\circ}$. 
The Moon shadow remains almost stable for the entire five-year observation period, which confirms stable operation of the detector. 
This result is consistent with the binned analysis in one dimension (Table \ref{tab:results_1d}).

The calculated center for the position of the Moon shadow is located in the middle of the Map, which also represents the expected position of the Moon. However, this method is only an estimator for the position of the Moon. 
A more sophisticated unbinned likelihood method to analyze the pointing accuracy of IceCube, given the known position of the Moon, is performed in \cite{Aartsen:2013zka}. 

\begin{figure*}[htbp]
	\centering
		\begin{tabular}{cc}
\includegraphics[trim = 5mm 0mm 5mm 0mm, clip, width=0.4\textwidth]{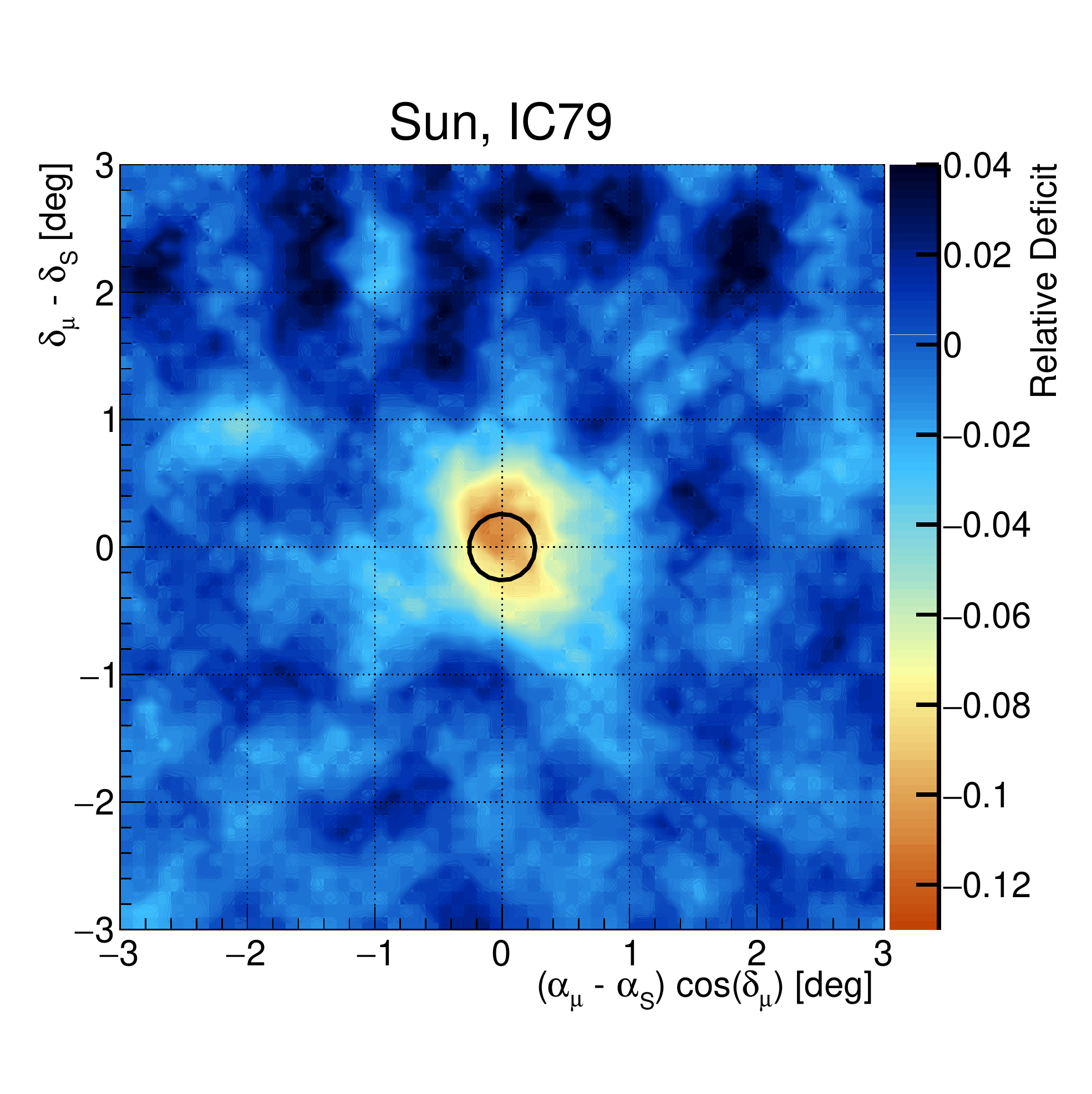}  & \includegraphics[trim = 5mm 0mm 5mm 0mm, clip, width=0.4\textwidth]{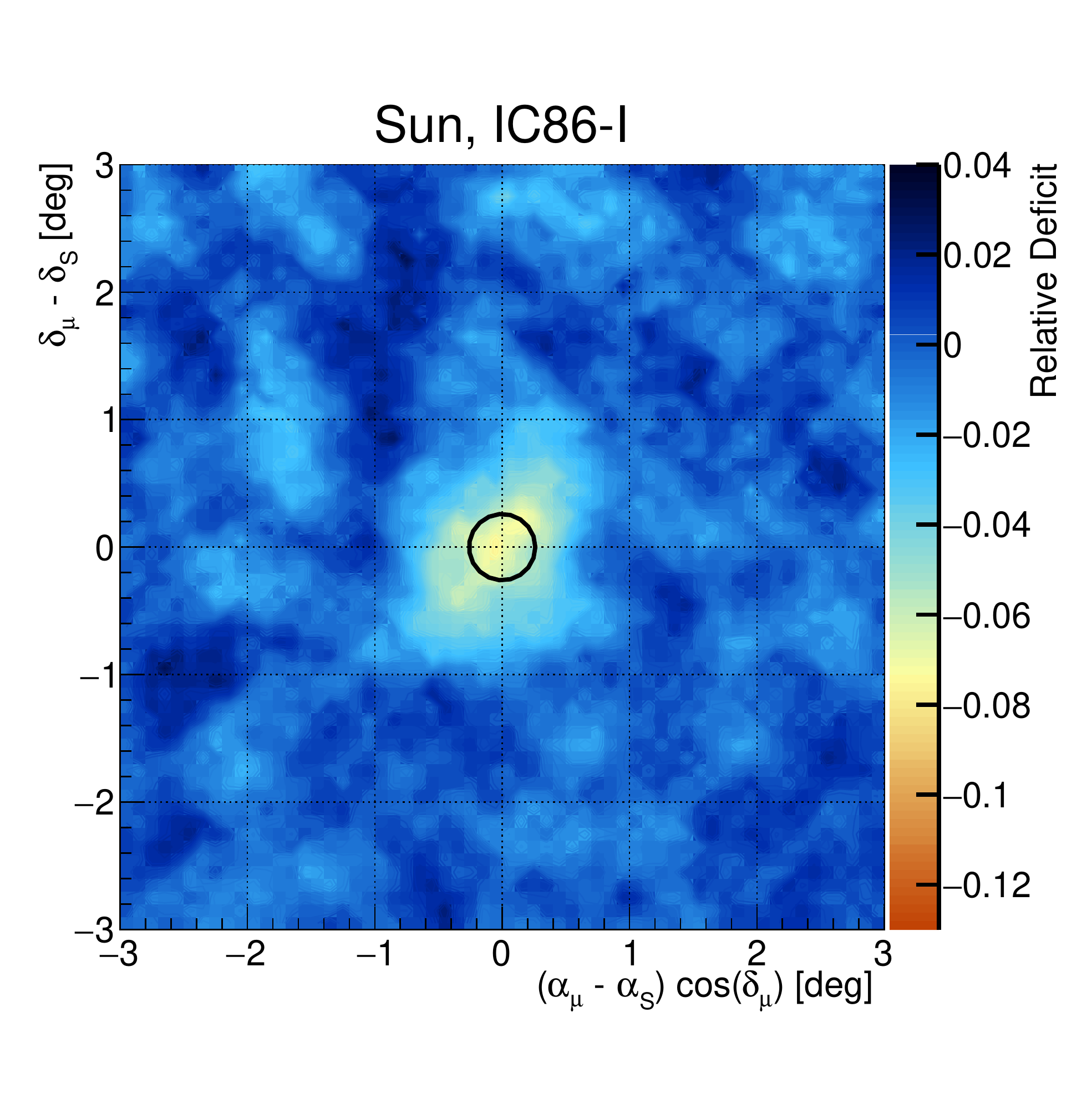} \\
\multicolumn{2}{c}{ \includegraphics[trim = 5mm 0mm 5mm 0mm, clip, width=0.4\textwidth]{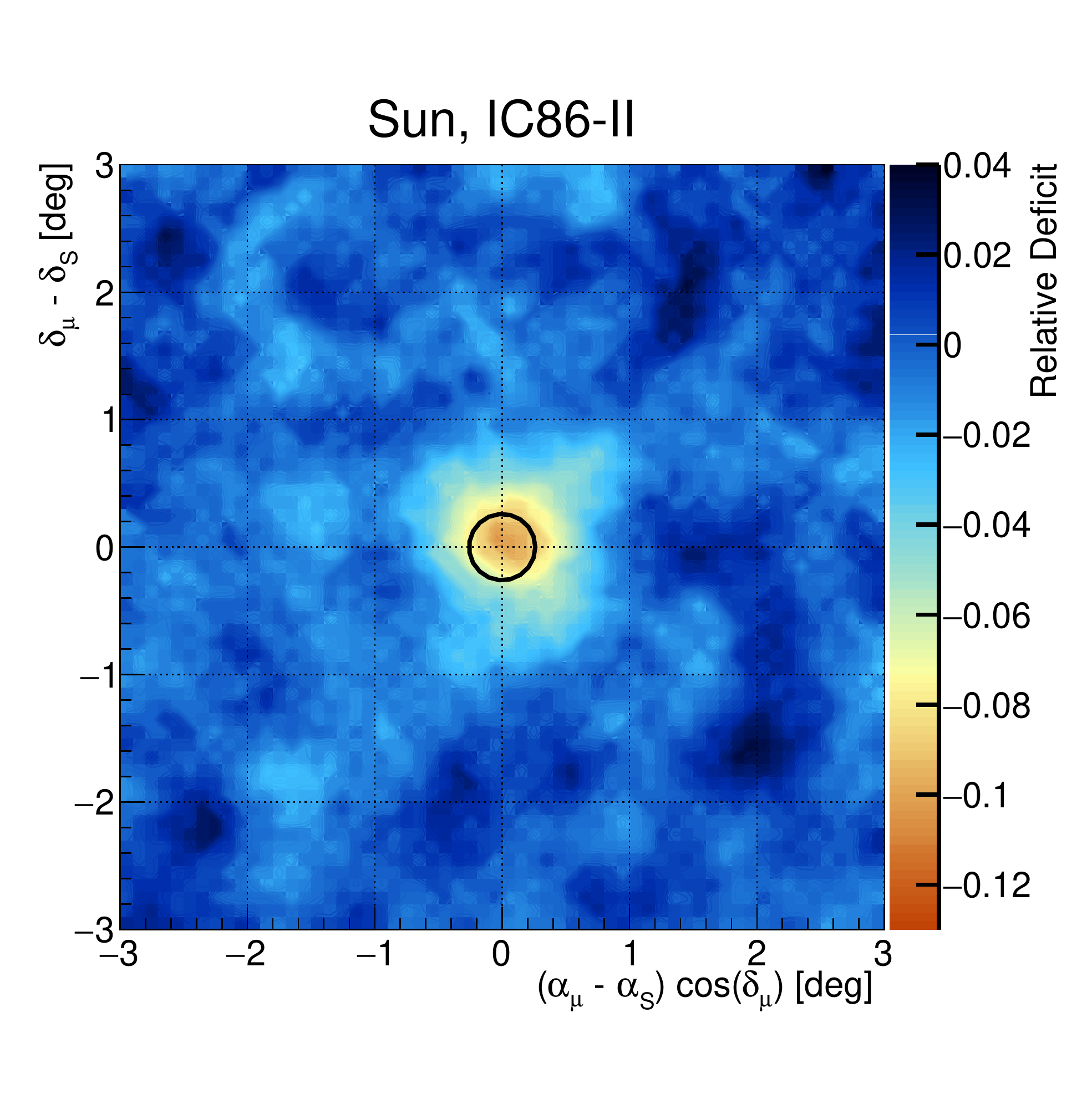} \hspace{0.23cm}  \includegraphics[trim = 5mm 0mm 5mm 0mm, clip, width=0.4\textwidth]{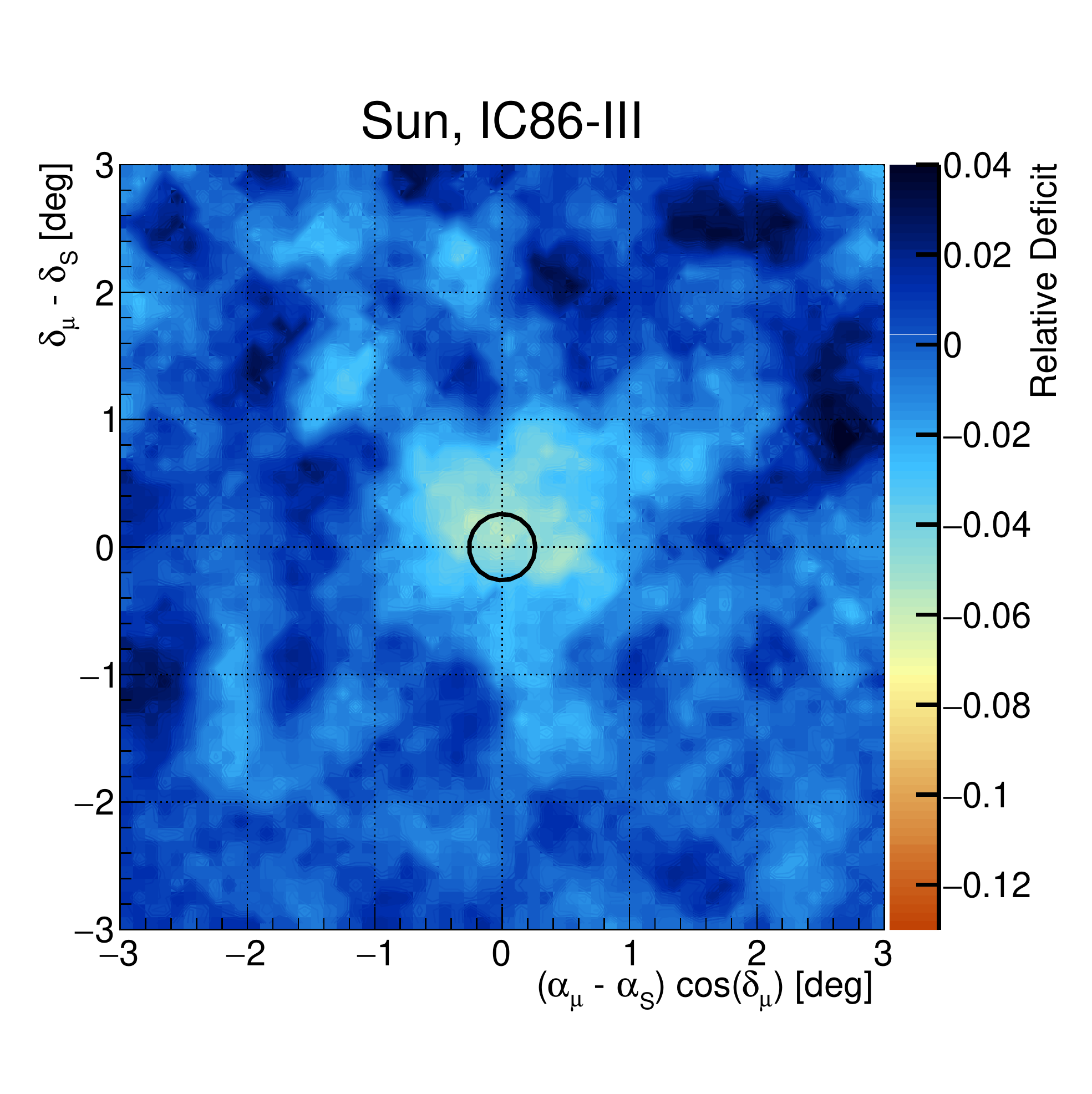}} \\
\multicolumn{2}{c}{ \includegraphics[trim = 5mm 0mm 5mm 0mm, clip, width=0.4\textwidth]{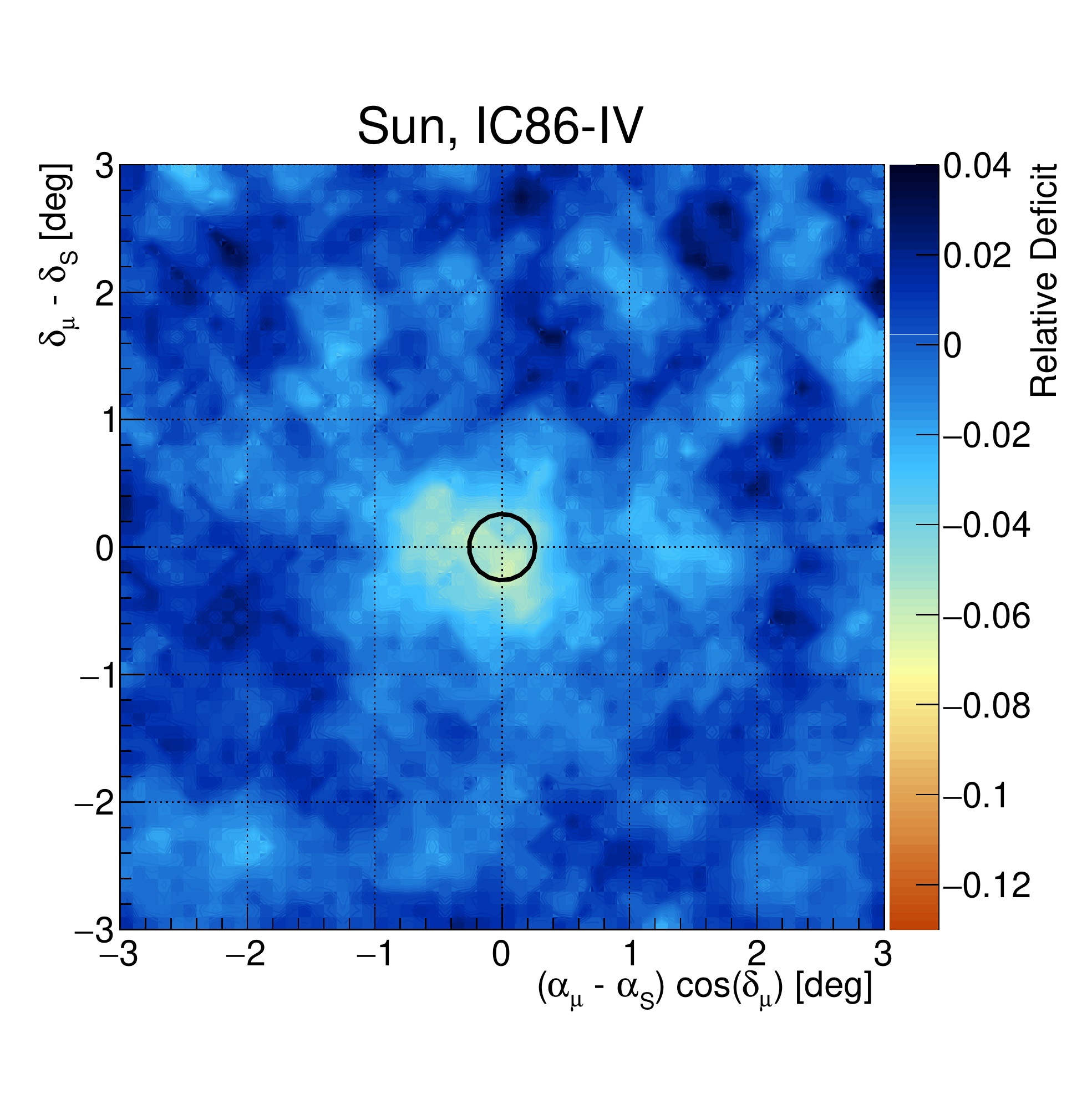}} \\
\end{tabular}
	\caption{Seasonal results for the 2-D binned maps of the Sun shadow. Each map uses data from November through February of each season. In IC79 the shadowing effect of the Sun is similar to the Moon shadow. Variations are clearly visible during IC86 I--IV (2011--2015) and are quantified in this paper (see text).}
	\label{fig:2d_sun}
\end{figure*}

The results for the two-dimensional binned approach of the Sun shadow can be seen in Figure \ref{fig:2d_sun}. 
Compared with the Moon shadow, the Sun shadow varies during the observation time, which confirms the results obtained in the 1-D analysis. 
A similar effect is observed in \cite{Zhu:2015ixa} for data from 2008 through 2012 and in \cite{2013PhRvL.111a1101A}. 
In \cite{2013PhRvL.111a1101A} it is shown that in the first year (2010/11) the shadowing effect is similar to the Moon shadow. 
In 2011/12 the shadowing effect decreases and increases in 2012/13. 
The weakest shadowing effect can be seen in 2013/14 and 2014/15. 
The next chapter will compare the measured Sun shadow with the solar activity. 

\clearpage
\section{Solar Activity}
\label{sec:solact}
The Sun has a magnetic field that varies over a 22 year cycle and displays magnetic activity in cycles of about 11 years. 
A direct estimator for the solar activity is the sunspot number. 
Cosmic ray particles are expected to be influenced by the solar magnetic field on their way to Earth. 
The binned analysis shows a constant Moon shadow and a time-variable Sun shadow. 
IceCube records data from the direction of the Sun from November through February each year. 
Due to the solar cycle, the solar activity changes for each season. 
A minimum of the monthly average sunspot number is observed with $24.5\pm 3.6$ in December 2010, a maximum with $146.1\pm 10.7$ in February 2014 by SILSO World Data Center (2016). 

\begin{figure*}[htbp]
\centering
\includegraphics[trim = 0mm 0mm 0mm 0mm, clip, width=0.74\textwidth]{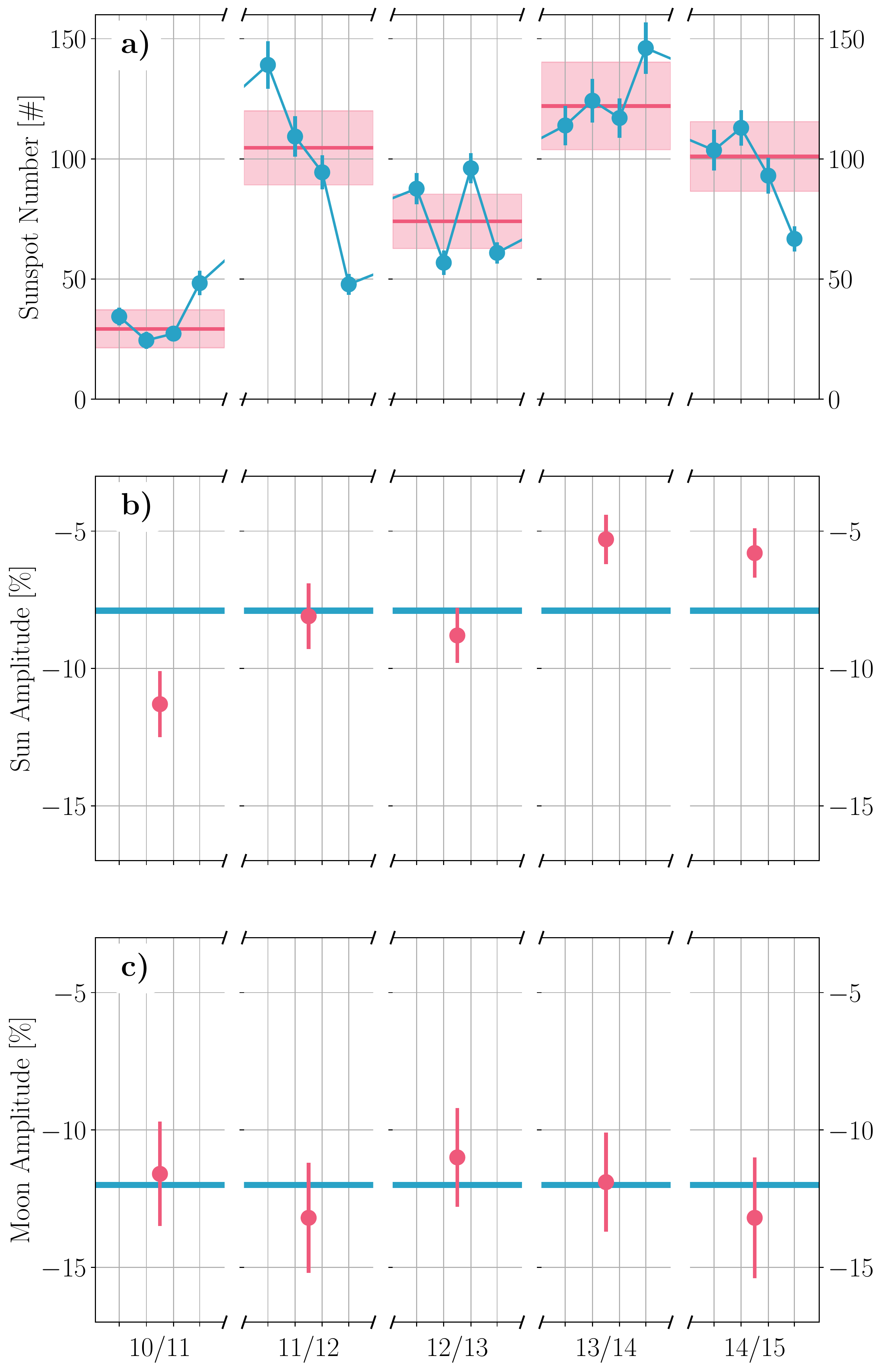}
\caption{Comparing the solar activity with IceCube's cosmic ray Moon and Sun shadow analysis. a.) Monthly mean sunspot number in November, December, January, February of each year. The red line shows the weighted mean of the sunspot number for each year. b.) Measured amplitude in the 1-Dimensional analysis for the Sun shadow using data. The blue line shows the mean amplitude across the five years of data. c.) Measured amplitude in the 1-Dimensional analysis for the Moon shadow using data. The blue line shows the mean amplitude across the five years of data without a modulation due to changes in the apparent size of the Moon which here amounts to about $\pm 0.5\%$. Sunspot numbers are taken from \cite{sidc}.}
\label{fig:sunspots}
\end{figure*}

\begin{figure*}[htbp]
\centering
\includegraphics[trim = 0mm 0mm 0mm 0mm, clip, width=0.74\textwidth]{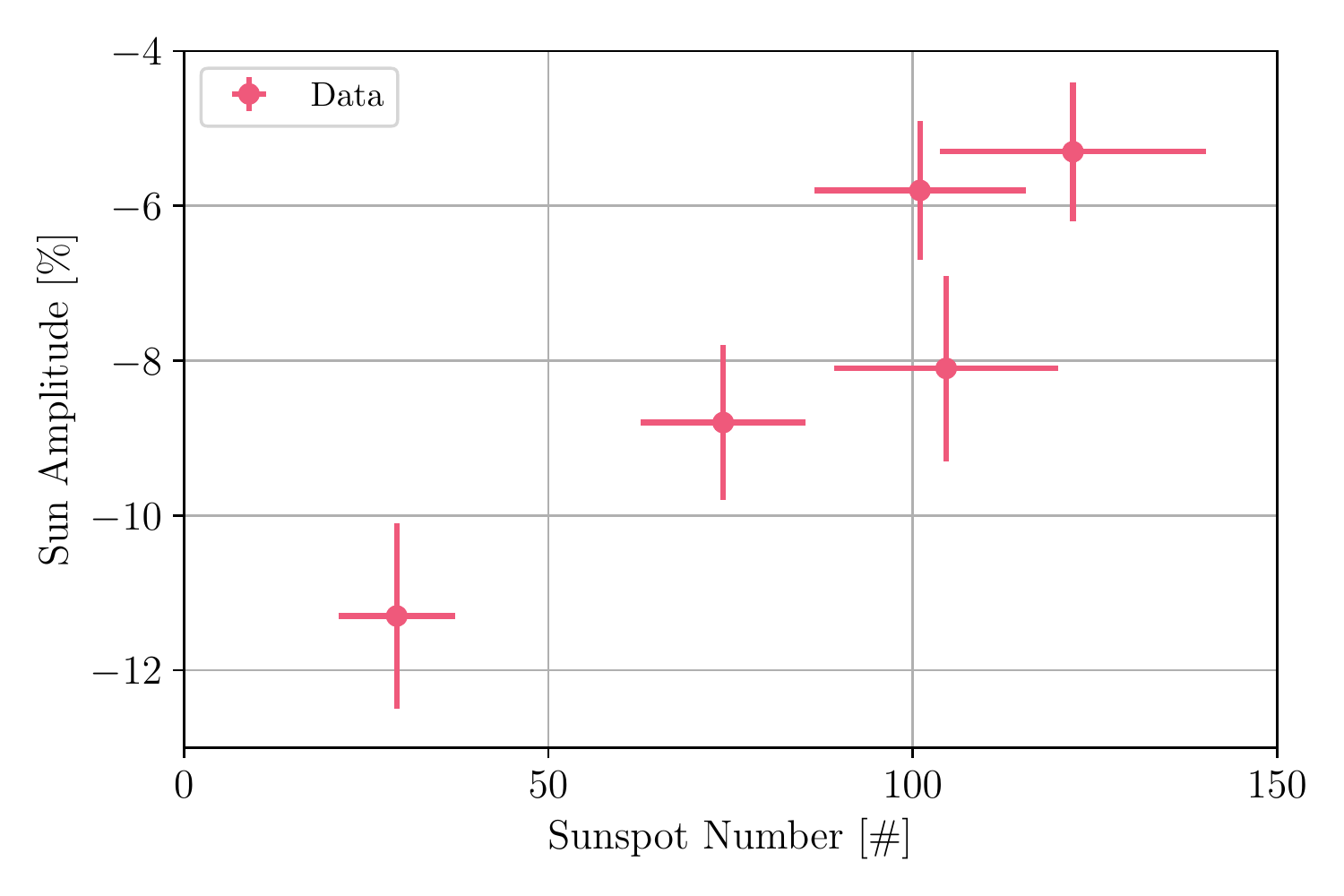}
\caption{Measured amplitude in the 1-Dimensional analysis for the Sun shadow as a function of the weighted average monthly sunspot number. A Spearman's rank correlation gives a hint to a correlation with a p-value of $96\,\%$. Sunspot numbers are taken from \cite{sidc}.}
\label{fig:sunspot_amplitude_correlation}
\end{figure*}

Figure \ref{fig:sunspots} shows a comparison of the amplitude of the fitted Gaussian in the binned 1-D analysis and the sunspot number from \cite{sidc}. 
The first plot (a) in Figure \ref{fig:sunspots} shows the weighted average sunspot number for each observation period, illustrated by the red line and the band as the statistical variation. 
Each monthly average sunspot number is weighted by the number of muon events in that particular month contributing to the final event sample of the corresponding observation period.
The monthly average sunspot numbers are represented by the blue data points. 
In (b) the amplitude of the fitted Gaussian of the Sun shadow analysis is shown by the red data points for each of the five detector configurations. 
The mean amplitude of the Sun shadow (blue line) is shown as a comparison. 
The mean amplitude of the Sun shadow data analysis is computed to $(7.9 \pm 0.5)\,\%$. 
The deviation of these five data points from the mean amplitude can be quantified by $\chi^2/\text{ndof} = 22.47/4$, which results in a statistical significance of $3.8 \sigma$. 
We conclude that there is a significant deviation from the mean shadowing effect of the Sun. 
A Spearman's rank correlation gives a hint to a correlation between the sunspot number and the amplitude of the Gaussian (compare figure \ref{fig:sunspot_amplitude_correlation}) with a p-value of $96\,\%$. 
However, this correlation needs to be studied further, due to a weak solar cycle and a small number of observed Sun periods at this point. 
The third plot (c) of Figure \ref{fig:sunspots} shows the amplitude of the Moon shadow analysis from Table \ref{tab:results_1d}. 
Here, the amplitude remains stable around its mean value $(12.2 \pm 0.9)\,\%$. 
These results show that the Moon shadow remains stable and can be used as a verification tool for the Sun shadow analysis.

\section{Conclusion}
The shadowing effects of Moon and Sun are observed with high statistical significance ($>10\sigma$) with the IceCube detector for every one-year data set, with data taken from its 79 and 86 strings configurations in a five-year observation period. 
These results can be compared with the Moon shadow analyses of previous detector configurations where the existence of the shadowing effect was shown at a level of ($>6\sigma$) significance \citep{Aartsen:2013zka}. 

This analysis, using 79 and 86 strings, respectively, achieves higher significance than the previous analysis, which had only 40 and 59 strings available. 
Using a binned analysis, a stable shadowing effect of the Moon is measured for the entire five-year period. 
This shows that the IceCube detector operates stably. 
The stable shadowing of the Moon also suggests that the magnetic fields between Earth and the Moon do not influence cosmic rays, at the energies of this analysis, significantly. 
However, the cosmic ray Sun shadow varies for each season. 

A significant difference between IceCube's measured Sun shadow and the mean shadowing effect of the Sun was determined as $ \chi^2 / \text{ndof}   = 22.47/4 $ resulting in a statistical significance of $3.8 \sigma$. 
This suggests that the shadow is not constant in time. 
An obvious explanation would be a variation with the Sun's magnetic field, as it was also detected during the previous solar cycle at lower energies by Tibet \citep{2013PhRvL.111a1101A}. 
A Spearman's rank correlation test shows that a correlation between the sunspot number and the amplitude of the Gaussian in the Sun shadow analysis is likely with $96\,\%$. 

However, due to the low number of observation periods in combination with a relatively weak solar cycle, this correlation must be investigated further. Additional observation seasons are necessary to verify the correlation between the solar activity and the shadowing effect of the Sun. 
Future analyses, including the deflection of cosmic rays due to magnetic fields between the Sun and Earth, should include a more complex treatment of the point spread function in order to compare results from data and simulations with respect to the integrated depth of the shadows.
\clearpage
\section*{Acknowledgements}
USA -- U.S. National Science Foundation-Office of Polar Programs,
U.S. National Science Foundation-Physics Division,
Wisconsin Alumni Research Foundation,
Center for High Throughput Computing (CHTC) at the University of Wisconsin-Madison,
Open Science Grid (OSG),
Extreme Science and Engineering Discovery Environment (XSEDE),
U.S. Department of Energy-National Energy Research Scientific Computing Center,
Particle astrophysics research computing center at the University of Maryland,
Institute for Cyber-Enabled Research at Michigan State University,
and Astroparticle physics computational facility at Marquette University;
Belgium -- Funds for Scientific Research (FRS-FNRS and FWO),
FWO Odysseus and Big Science programmes,
and Belgian Federal Science Policy Office (Belspo);
Germany -- Bundesministerium f\"ur Bildung und Forschung (BMBF),
Deutsche Forschungsgemeinschaft (DFG),
Helmholtz Alliance for Astroparticle Physics (HAP),
Initiative and Networking Fund of the Helmholtz Association,
Deutsches Elektronen Synchrotron (DESY),
and High Performance Computing cluster of the RWTH Aachen;
Sweden -- Swedish Research Council,
Swedish Polar Research Secretariat,
Swedish National Infrastructure for Computing (SNIC),
and Knut and Alice Wallenberg Foundation;
Australia -- Australian Research Council;
Canada -- Natural Sciences and Engineering Research Council of Canada,
Calcul Qu\'ebec, Compute Ontario, Canada Foundation for Innovation, WestGrid, and Compute Canada;
Denmark -- Villum Fonden, Danish National Research Foundation (DNRF), Carlsberg Foundation;
New Zealand -- Marsden Fund;
Japan -- Japan Society for Promotion of Science (JSPS)
and Institute for Global Prominent Research (IGPR) of Chiba University;
Korea -- National Research Foundation of Korea (NRF);
Switzerland -- Swiss National Science Foundation (SNSF).
The IceCube collaboration acknowledges the significant contributions to this manuscript from Fabian Bos and Frederik Tenholt.
\pagestyle{empty}
\bibliographystyle{aasjournal}
\bibliography{lit}
\end{document}